\documentclass[sn-mathphys,Numbered]{sn-jnl}

\usepackage{graphicx}%
\usepackage{multirow}%
\usepackage{amsmath,amssymb,amsfonts}%
\usepackage{amsthm}%
\usepackage{mathrsfs}%
\usepackage[title]{appendix}%
\usepackage{xcolor}%
\usepackage{textcomp}%
\usepackage{manyfoot}%
\usepackage{booktabs}%
\usepackage{algorithm}%
\usepackage{algorithmicx}%
\usepackage{algpseudocode}%
\usepackage{listings}%
\usepackage{bm}
\usepackage{bbold}

\newcommand{\ket}[1]{\left|#1\right\rangle}

\newcommand{\bra}[1]{\left\langle#1\right|}
\newcommand{\braket}[2]{\langle#1|#2\rangle}

\newcommand{\ketbra}[2]{|#1\rangle\langle#2|}
\newcommand{\dketbra}[1]{|#1\rangle\langle#1|}
\newcommand{\be}{\begin{equation}}

\newcommand{\ee}{\end{equation}}

\newcommand{\cB}{{\cal B}}

\newcommand{\cC}{{\cal C}}

\newcommand{\cH}{{\cal H}}

\DeclareMathOperator*{\argmax}{arg\,max}

\newcommand{\bbb}{{\bf b}}
\newcommand{\bc}{{\bf c}}

\newcommand{\btheta}{{\boldsymbol \theta}}
\newcommand{\balpha}{{\boldsymbol \alpha}}

\newcommand{\tr}[1]{{\rm Tr}\left[#1\right]}

\renewcommand{\i}{{\rm i}}
\newcommand{\one}{\mathbb{1}}

\begin{document}

\title[]{Optical decoder learning
for fiber communication at the quantum limit}

\author*[1]{\fnm{Matteo} \sur{Rosati}}\email{matteo.rosati@uniroma3.it}

\author[2]{\fnm{Albert} \sur{Solana}}


\affil[1]{\orgdiv{Dipartimento di Ingegneria Civile, Informatica e delle Tecnologie Aeronautiche}, \orgname{Università Roma Tre}, \orgaddress{\street{Via Vito Volterra 62}, \city{Rome}, \postcode{I-00146}, \country{Italy}}}

\affil[2]{\orgname{Qilimanjaro Quantum Tech}, \orgaddress{\street{Passeig de Gràcia 58}, \city{Barcelona}, \postcode{08007}, \country{Spain}}}

\abstract{Quantum information theory predicts that communication technology can be enhanced by using quantum signals to transfer classical bits. 
In order to fulfill this promise, the message-carrying signals must interact coherently at the decoding stage via a joint-detection receiver (JDR). 

To date, the realization of a JDR using optical technologies has remained elusive: the only known explicit design is the Hadamard receiver, which increases distinguishability at the cost of reducing the code size. Therefore, the discovery of efficient and scalable JDR designs is an outstanding open problem for the demonstration of a quantum advantage in fiber and space communication.

We introduce a supervised-learning framework for the systematic discovery of new JDR designs based on  parametrized photonic integrated circuits. Our framework relies on the synthesis of a training set comprising quantum codewords and the corresponding classical message label; the codewords are processed by the JDR circuit and, after photo-detection, produce a guess for the label. The circuit parameters are then updated by minimizing a suitable loss function, reaching an optimal JDR design for that specific architecture. 

We showcase our method with coherent-state codes for the pure-loss bosonic channel, a paradigmatic model of optical-fiber and space communication, with a circuit architecture comprising the most general linear-optical gate, squeezing and threshold photo-detectors. We train JDR circuits for several families of classical codes, i.e., random, linear and polar, varying the mean signal energy and the message size.

We discover optical JDR circuit setups for maximum-size codes and small message-length that offer up to a $3$-fold enhancement in the bit decoding rate with respect to the optimal single-symbol receiver, and less than $7\%$-away from the theoretically optimal decoder, for which an explicit design is missing to date. Furthermore, the discovered receivers surpass the performance of the Hadamard receiver both in terms of bit decoding and bit transmission rate.

Finally, we interpret the discovered setups to gain analytical insight for future code design: the best-performing codes are those which can be mapped, via the JDR's optical processing, to modulations with different energy levels on different symbols, making the message symbols more distinguishabile via photo-detection. 
}




\maketitle

\section{Introduction}
\label{sec1}

The digital revolution has reshaped society during the last thirty years, marking the advent of the information age and the permeation of information technology in every aspect of our lives. The capability of exchanging messages between distant users forms the backbone of this revolution, and the stepping stone for future technological advancements. As the volume of transmitted information grows closer to the channel capacity~\cite{Bayvel2016}, i.e., the maximum bit transmission rate allowed by classical physics, and non-linear effects in fiber come under consideration~\cite{Mitra2001,Essiambre2010}, technology challenges the ultimate limits of nature, calling upon a quantum-mechanical description of information processing. 

In this setting, quantum information theory has long predicted that the Shannon capacity can be surpassed by harnessing the quantum properties of light, achieving the Holevo capacity~\cite{helstromBOOK,holevoBOOK,Hausladen1996,Schumacher97,holevo1998c,winter,hayanaga,hastings2009,wildeBOOK}. This achievement, in turn, would have numerous practical implications, including: efficient communication in photon-starved scenarios, e.g., long-distance and deep-space communication
~\cite{Shapiro2005,Waseda10,Waseda11,Banaszek2012,Powell2013,Jarzyna2015,Jarzyna2019a,Jarzyna2019b,Banaszek2020}, a $57\%$ reduction of energy consumption in medium-distance scenarios~\cite{Notzel2022}, e.g., optical fiber networks, as well as the potential of enhancing other quantum communication services, e.g., quantum key distribution and entanglement-assisted communication~\cite{Guha2020}.%

Unfortunately, much like the Shannon capacity, the mere existence of a Holevo-capacity-achieving protocol in theory has not granted the realization of a practical algorithm to date. Indeed, it is well-known that a simple encoding via product sequences of optical coherent-states suffices to attain the optimal communication rate for a large class of Gaussian channels, modelling the effects of loss, noise and amplification in fiber and space links~\cite{Giovannetti2004,Giovannetti2013a,gaussMaj,DePalma2014,DePalma2017}. However, the decoder is required to perform complex multi-symbol measurements, implementing a quantum-coherent interaction between successive signals, known as joint-detection receiver (JDR)~\cite{Schumacher97,holevo1998c,guha2,Guha2010,Guha11,Giovannetti2011a,Giovannetti2012,guha2012,Chen2012,seqCoh,Takeoka14,Klimek2015,Klimek2015a,Rosati16b,Rosati16c,Rosati2017,Rengaswamy2020}, which harnesses the fundamental effect of non-locality without entanglement~\cite{Bennet1999}. Such challenging task is further aggravated by our limited technological capabilities: the efficient manipulation of continuous-variable states of light is at present still restricted to so-called Gaussian operations, i.e., linear interferometers and squeezing, plus a limited class of non-Gaussian resources, mainly photo-detectors~\cite{Zhong2020,Madsen2022}.  

In light of the afore-mentioned reasons, the realization of a JDR with current photonic technology remains an outstanding open problem to date. The only known explicit decoder design, proposed by Guha~\cite{Guha11} (see also~\cite{Klimek2015,Rosati16c}), is the Hadamard receiver (HR), which enhances codeword distinguishability at the cost of sacrificing the code size. Thanks to these properties, the HR is able to surpass the Shannon limit in long-distance unamplified communication scenarios where the received signals carry much less than one photon on average. Still, the HR remains unable to achieve the Holevo capacity in such scenarios; furthermore, its performance quickly falls below the Shannon limit for larger signal power~\cite{Guha11,Rosati16c,Banaszek2020}, motivating the search for new receiver designs that work closer to the optimal and for larger photon numbers. \\

\begin{figure}[h!]
{\centering\includegraphics[width=\textwidth,trim={0 6cm 0 5cm},clip]{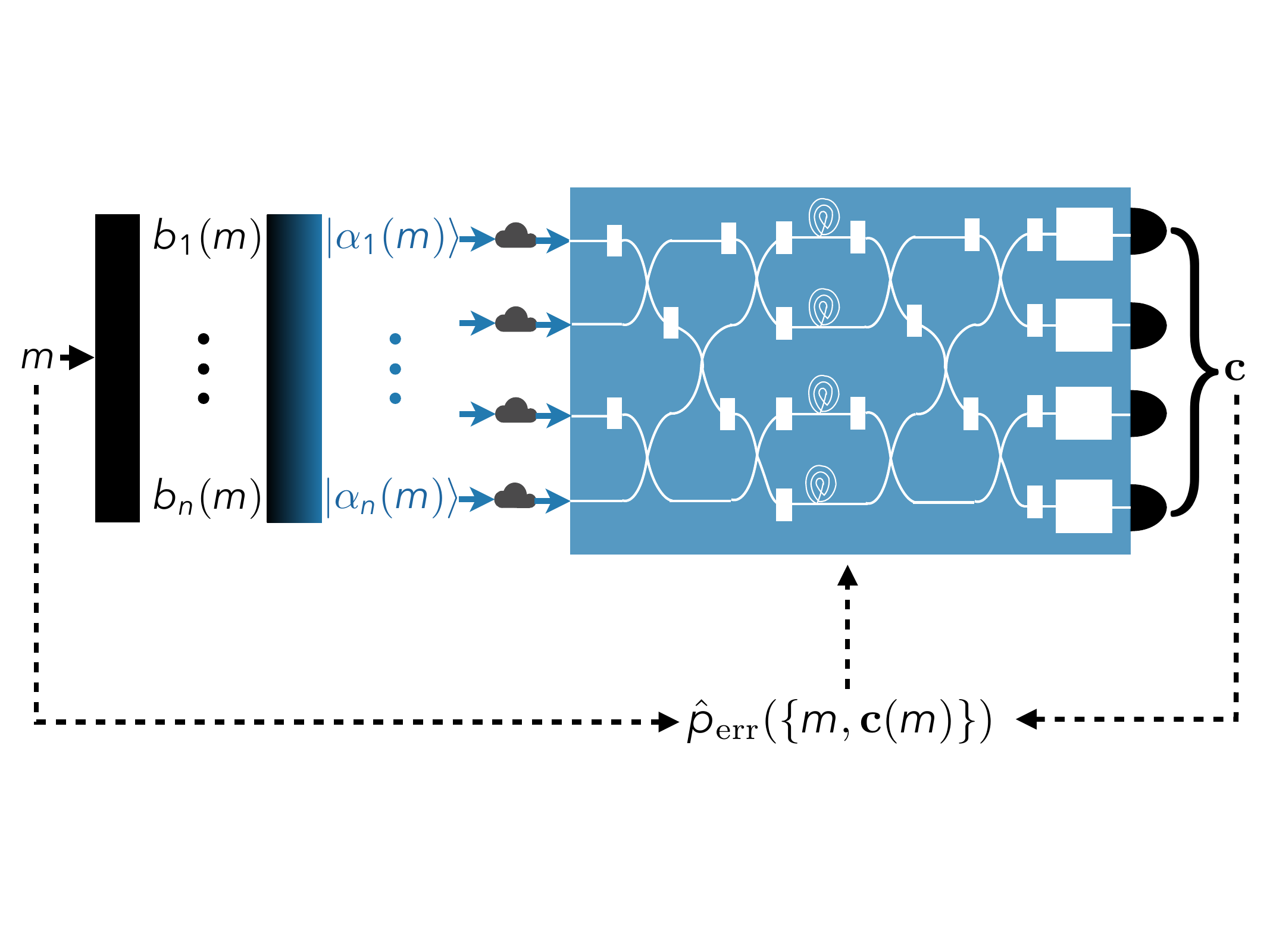}}
\caption{Schematic depiction of our training algorithm applied to the discovery of optical decoders for quantum-enhanced bit transmission: an input message $m$ is encoded via a classical code into a bit string $\bbb(m)$; the latter is then encoded into a product-state sequence of optical coherent states $\ket{\balpha(m)}$, e.g., via a binary-phase modulation with phases $(-1)^{b_i(m)}$ on mode $i$ (see text). After transmission on the channel, the received quantum codeword is processed via a general Gaussian optical circuit including linear-optical interferometers, displacements and squeezers, implementable with integrated photonics. The circuit gates are characterized by a set of parameters $\btheta$ and they are reconfigurable. The circuit output is measured via threshold photodetectors, providing a binary outcome string $\bc$. By repeating the experiment with a fixed circuit setup $\btheta$ and a batch of random codewords, one can estimate the average decoding error probability (see text) and propose a new circuit setup $\btheta'$ that improves the error performance. The training is repeated batch-by-batch until reaching a desired performance.}\label{fig:scheme}
\end{figure}

Here we introduce a supervised-learning framework that is capable of learning a near-optimal receiver configuration within a parametrized photonic integrated circuit, given a coherent-state code as input (see Fig.~\ref{fig:scheme}). In recent years, the introduction of machine learning techniques has increased our capabilities of searching large parameter spaces, with successful applications in the calibration and control of physical and virtual quantum hardware, such as photonic integrated devices~\cite{Schuld2018,Arrazola2018,Killoran2018,Killoran2018b,Steinbrecher2019,Bilkis2020,Bilkis2021,Saggio2021}. For the first time to our knowledge, we harness such capability for the design of quantum-to-classical decoders: we generate a training set of random codewords with corresponding message labels; the former are sent through the channel and then detected via the parametrized circuit, which makes a guess for the message. This is used, together with the true label, to compute a suitable loss function and update the circuit parameters for a new run. The training procedure is repeated until the score function does not further improve.

Using our framework, we are able to discover a variety of near-optimal explicit receiver designs for families of linear, random and polar codes~\cite{gallagerBOOK,Arikan2009,Wilde2013} with small block-lengths, up to $6$ modes, and maximum code size. At variance with previous single-mode results~\cite{Takeoka2008,Sidhu2021}, we show that the use of ancillary modes and/or squeezing during detection does not significantly increase the decoder performance. Strikingly, we identify receiver designs provably JDR and with record performance, exhibiting a bit-decoding rate up to $3$ times higher than the best single-symbol receiver, and only $7\%$-away from the optimal one. Importantly, our receiver designs surpass, for the first time to our knowledge, the net bit decoding and transmission rate of the HR with the same number of modes. Furthermore, we show that our decoders also outperform the homodyne receiver and HR in terms of practically relevant features: (i) they are also more energy- and space-efficient; and (ii) they deliver a clear advantage for communication at finite-blocklength, thus requiring a moderate effort to prove the enhancement of non-locality without entanglement in a practical bit-transmission protocol.

Finally, we study the best code-decoder couples found, to provide new insights for the development of optimal decoders suitable for larger message-lengths. Remarkably, our decoders depart from the \emph{one-signal-per-mode} HR paradigm, whose efficiency relies on assigning to each codeword a single pulse in distinct spatial or temporal modes. As a matter of fact, we show that larger and more space-efficient codes can be efficiently decoded by assigning to each codeword a distinct pulse-energy profile along the available modes. Such profiles are still easily distinguishable via threshold-photodetection, and we conjecture that they are optimal  at even larger signal energies.\\

The application of our framework to the simplest non-trivial message-length provably unlocks interesting decoder designs with excellent performance. Additionally, our framework marks a starting point, easily extendable to include many desirable features and tackle larger sizes, e.g.: (i) the use of further non-Gaussian circuit components that may become available in the future; (ii) the introduction of circuit imperfections, countering the decrease in performance via a different parameter choice; (iii) the use of adaptive-feedback loops, which are known to further enhance performance; (iv) training on an actual physical circuit, rather than a simulated one, thus automatically including imperfections, increasing the overall training speed, and widening the accessible range of energies and block-lengths. 
 
In light of these reasons, the introduction of a systematic framework for the discovery of quantum-to-classical decoder designs sets an example for future endeavours, opening a new direction for the development of Holevo-capacity-achieving protocols with photonic devices, that will be easily integrated in the current communication infrastructure, enhancing their performance and diversifying their capabilities.

\section{Results}\label{sec:results}

\subsection{Bit transmission at the quantum limit: challenges and promises}
We consider the typical quantum model of bit transmission using optical signals that travel in fiber or free-space. The sender encodes her message in a sequence of quantum states of the electromagnetic field, or ``quantum letters"; each letter belongs to an infinite-dimensional Hilbert space $\cH_i$ of a distinct field mode, e.g., separated in time, space or frequency from the other modes; each mode is represented by a couple of photon-creation and -annihilation operators $ a_i,  a_i^\dag$, satisfying the canonical commutation relations $[ a_i,  a_i^\dag]=1$\footnote{Here and in the rest of the article we set $\hbar=1$ and restrict to temporally- or spatially-separated narrow-band signals with unit frequency $\omega=1$, without loss of generality. However, our results also apply to frequency-separated signals travelling in parallel on a certain bandwidth, by assigning different frequencies $\omega_i$ to each $a_i$. }. Transmission losses are modelled as a pure-loss bosonic Gaussian channel mapping $ a_i \mapsto \sqrt{\eta}\,  a_i + \sqrt{1-\eta}\, e_i$, where $ e_i$ describes an environmental mode that we assume to be in the vacuum state (no added noise except for vacuum fluctuations\footnote{With this choice, the added noise is $\frac{1}{2}$ for each field quadrature in shot-noise units. Extra channel noise can be included in the model and, if significantly large, it can affect the performance of coherent-state decoders~\cite{}. Nevertheless, the methods we develop can be applied also to this case to find an optimal decoder design under larger-noise conditions}). The receiver can decode the message by performing a general quantum measurement on the collective Hilbert space $\cH = \bigotimes_i \cH_i$ of the letters' sequence, or ``quantum codeword".

In a seminal result, Holevo~\cite{holevoBOOK,holevo1998c} showed that, if one allows for arbitrary encoding and decoding operations, the maximum bit transmission rate is
\begin{equation}\label{eq:holevo_capacity}
    \chi(E) = (E + 1) \log(E + 1) - E \log(E),
\end{equation}
where $E=\eta\,E_0$ is the mean number of \emph{received} photons of the quantum codewords and $E_0$ the corresponding quantity at the channel input\footnote{Here and in the rest of the article we use base-$2$ logarithms, measuring rates in number of bits per channel use.}. 
Subsequently, for a wide class of channels including loss, noise and amplification, it was proved that the use of optical coherent states as quantum letters can attain the Holevo capacity~\cite{Giovannetti2004,Giovannetti2013a,gaussMaj,DePalma2014,DePalma2017}; such property incredibly simplifies the encoding procedure, since the message can be encoded bit-by-bit into a product quantum state without the need to generate entanglement. However, the same cannot be said of the receiver: to date, it is believed that a collective (non-product) quantum measurement, is necessary to optimally decode the messages and achieve the Holevo capacity; such a decoder, that performs coherent quantum interactions between the quantum letters, is called a joint-detection receiver (JDR)~\cite{Schumacher97,holevo1998c,guha2,Guha2010,Guha11,Giovannetti2011a,Giovannetti2012,guha2012,Chen2012,seqCoh,Takeoka14,Klimek2015,Klimek2015a,Rosati16b,Rosati16c,Rosati2017,Rengaswamy2020}. 

In contrast, if one allows only single-mode detection, performing quantum measurements to decode the letters one by one, the bit transmission rate is bounded by the Shannon capacity of the classical channel induced by the measurement. As a benchmark at small $E$, one can consider the classical strategy of homodyne detection~\cite{Banaszek2020}:
\begin{align}\label{eq:hetero_homo_capacity}
    C_{\rm homo}(E) = \frac12 \log(1 + 4E).
\end{align}
The significant gap between \eqref{eq:holevo_capacity} and \eqref{eq:hetero_homo_capacity} motivates the intense on-going research for an explicit JDR design that can attain the Holevo capacity. The potential for applications is remarkable, ranging from a net increase of the bit transmission rate in unamplified long-distance links, e.g., space communication~\cite{Waseda10,Waseda11,Banaszek2020}, which could benefit as a by-product also quantum key distribution and entanglement-assisted communication protocols~\cite{Guha2020}, to a $57\%$ decrease of energy consumption from optical amplification in fiber networks~\cite{Notzel2022}.

The main challenge is finding an optimal JDR implementation within a restricted class of quantum circuits which can be implemented using integrated photonics~\cite{Wang2020,Pelucchi2021,Madsen2022}, comprising~\cite{Takeoka2008,Takeoka14,Rosati2017}: (i) Gaussian operations (G), i.e., interferometers and squeezing; (ii) photodetection measurements (P); and (iii) classical adaptive control (C); we refer to this class as GPC. To date, the Hadamard receiver (HR)~\cite{Guha11,Klimek2015,Rosati16c} is a unique example of JDR that surpasses single-symbol receivers; unfortunately, its advantage is limited to the extreme long-distance regime $E\lesssim 10^{-1}$ and it is unable to fully close the gap with the Holevo capacity~\cite{Rosati16c} (see Methods~\ref{subsec:theory_competitors}). Therefore, we are still lacking a method for designing receivers within the GPC class that can close the gap with the Holevo capacity by increasing the codewords' length, or block-length, for a wide range of received photon numbers. In the next subsections, we present a supervised learning framework that can tackle this problem and show its first promising results, introducing receiver designs that surpass the HR.

\subsection{Supervised decoder learning}
Our method requires as input a code $\cC$, i.e., the set of quantum codewords that will be used to transmit different messages. For simplicity, we restrict to binary-phase coherent-state codes such that the \emph{received} codewords are products of coherent states $\ket{\pm \alpha}$ with $\alpha = \sqrt{E}$\footnote{A single-mode coherent state is the eigenstate of the photon-annihilation operator, $ a_i \ket{\alpha} = \alpha \ket{\alpha}$}. These codes, albeit not sufficient to attain the Holevo capacity in general, are well-performing at low mean photon number (see Methods~\ref{subsec:theory_competitors}). Furthermore, an understanding of optimal decoding for binary codes is the foundation for the design of decoders adapted to larger modulations, e.g., by combining a binary code with a multi-phase modulation~\cite{Rosati16c}. A good binary-phase quantum code can be generated by mapping first the input message $m\in\{1,\cdots,|\cC|\}$ into a classical bit-string $\bbb(m)\in\{0,1\}^n$, whose bits are then used to modulate the signs of a product of coherent states
\begin{equation}
    \ket{\balpha(m)} = \ket{\alpha_1(m)}_1 \otimes\cdots\otimes\ket{\alpha_n(m)}_n,
\end{equation}
with $\alpha_i(m)=(-1)^{b_i(m)} \sqrt{E_0}$ (see Fig.~\ref{fig:scheme}). The map $m\mapsto \bbb(m)$ is provided by a capacity-achieving classical binary code; see Methods~\ref{subsec:codes} for a detailed discussion of the linear, random, and polar codes that we employ. After transmission, the codewords present the same sign modulation but with smaller amplitude: $\alpha_i(m)=(-1)^{b_i(m)} \sqrt{E}$.

The receiver consists of an $n$-mode Gaussian unitary $U(\btheta)$, which can be decomposed into basic gate components, i.e., beam splitters, phase-shifters, squeezers and optical displacements, completely determined by a vector of $O(n^2)$ real parameters $\btheta$, followed by threshold photodetectors\footnote{Photon-counting detectors can be employed as well, with potentially better performance. Here we restrict to simple threshold detectors in order to decrease the computational burden of simulating the quantum device, which can be lifted by increased computational resources or the use of a physical device.} (see Methods~\ref{subsec:circuit}). In this case, the measurement outcomes are also bit-strings $\bc\in\{0,1\}^n$, with conditional probability given by the Born rule:
\begin{equation}\label{eq:output_probabilities}
    P(\bc | m) = \bra{\balpha(m)}U(\btheta)^\dag M_{\bc} U(\btheta)\ket{\balpha(m)},
\end{equation}
where 
\begin{equation}\label{eq:measurement_ops}
    M_{\bc} = \bigotimes_{i=1}^n \left(\dketbra{0}^{c_i}\left(\one-\dketbra{0}\right)^{1-c_i}\right)
\end{equation}
corresponds to a product of projectors on vacuum-states ($c_i=0$) vs. the subspace with one or more photons ($c_i=1$).
Based on the outcome, the decoder makes a maximum-likelihood guess, choosing the most likely input given the outcome $\bc$:
\begin{equation}\label{eq:max_likelihood}
   \hat m(\bc) =  \argmax_{m} P(\bc | m),
\end{equation}
since the codewords are taken equiprobable. Note that this receiver falls within the GP class, since we do not allow partial signal measurements and adaptive control based on the measurement outcomes. 

The principal figure of merit to evaluate a decoder's performance in practical terms is the average probability of error in decoding the message:
\begin{equation}\label{eq:loss}
    p_{\rm err}(\cC,\btheta) = 1-\frac1{|\cC|}\sum_{\bc}  \max_{m} P(\bc | m).
\end{equation}
Indeed, for a capacity-achieving protocol it holds that, given a sequence of codes of size $|\cC_n| = \lfloor 2^{n R} \rfloor$, with $\chi(E) - \epsilon< R < \chi(E)$ and any $\epsilon>0$, the average probability of error of the decoder approaches zero as the message length $n$ tends to infinity. Hence, for a given received number of photons $E$ and message length $n$, we employ codes of capacity-achieving size and score each decoder parameter setup with the function \eqref{eq:loss}. 

The quantum circuit $U(\btheta)$ can then be regarded as a single-layer continuous-variable quantum neural network~\cite{Killoran2018}, and trained to classify the input codewords by assigning to them an estimate $\hat m(\bc)$ of the true message label depending on the measurement outcome $\bc$. Here, the main difficulty with respect to other applications is to devise an algorithm that takes into account the maximum-likelihood guessing rule; as a matter of fact, such rule can be applied for a given outcome $\bc$ only when the probabilities $P(\bc |m)$ are known for all possible messages $m$, or at least when they can be approximated to a certain extent. The data presented in this article were produced using a virtual device built with Xanadu's Strawberryfields simulator~\cite{Killoran2018b}, where one can obtain the outcome probability directly from the circuit parameters $\btheta$ in a single shot. This function can thus be used as an oracle to implement our probability-based supervised-learning algorithm, described as follows:
\begin{enumerate}
    \item Construct a training batch $\cB$ of random codewords sampled without replacement with equal probability from $\cC$, and their corresponding classical message label,
    \begin{equation}
    \cB=\{(|\balpha(m_j)\rangle,m_j)\}_{j=1}^{|\cB|};
    \end{equation}
    \item For each codeword $\ket{\balpha(m_j)}$ and fixed parameters $\btheta$, run the virtual device $U(\btheta)$ with that codeword as input and obtain the outcome probabilities $P(\bc|m)$ for all $\bc$ from the oracle;
    \item After analyzing all the codewords in the batch, for each possible outcome compute the maximum-likelihood guess as \eqref{eq:max_likelihood};
    \item Compute an estimate of the loss function on the current batch:
        \begin{equation}
            \hat p_{\rm err}(\cB,\btheta) = 1-\frac1{|\cB|}\sum_{\bc} P(\bc| \hat m(\bc));
        \end{equation}
    \item  Obtain a new candidate set of parameters $\btheta'$ for the next iteration via a function-optimization method (see Methods~\ref{subsec:algorithms}).
\end{enumerate}
Iterating this procedure on a suitable number of batches, or until the loss function does not show significant improvements by further iterations, produces a final candidate set of parameters $\btheta_{\rm op}$ that fully describe the near-optimal receiver circuit found, in terms of the unitary $U(\btheta_{\rm op})$. 

If a virtual oracle is not available, then one has to estimate the probabilities $P(\bc|m)$ directly from multiple runs of a physical device, as detailed in Methods~\ref{subsec:algorithms}. We expect that running on a physical device will reduce the training time; indeed, the single runs will be faster, not requiring a simulation of the circuit, and furthermore a high-probability subset of outcomes for any given codeword will suffice to discard bad decoder setups. 

We observe that a similar supervised-learning approach to decoder design was recently investigated in the classical setting~\cite{Nachmani2016,Gruber2017,Cammerer2017,Weinberger,Kim}, with significant results compared to state-of-the-art classical decoders. The added difficulty in our setting is that the decoding neural network consists of quantum gates, which are more demanding to simulate, whose training is not as well understood as for classical neural network, and, for our particular case, which consist of a non-universal set of gates. Furthermore, in the classical approaches the output of the decoding network is deterministic, unlike ours. 

\subsection{Optimal decoder analysis}
We run our algorithm using as input codebooks constructed from linear, random and polar codes, for a variety of mean-photon-number values $E$ and number of message modes $n$. In the following we provide a comprehensive analysis of the performance of the discovered decoders, proving that the outperform the state-of-the-art.\\

We start by analyzing the performance of decoders for linear codes based on linear-optical circuits, i.e., no squeezing, and a number of decoder modes equal to the number of message modes $n$, i.e., no ancillae. We compare our decoders with the best-performing competitors in the literature, and in terms of several performance metrics; see Methods~\ref{subsec:theory_competitors} for further details and explicit expressions. 

\begin{figure}[h!]
{\centering\includegraphics[scale=.32]{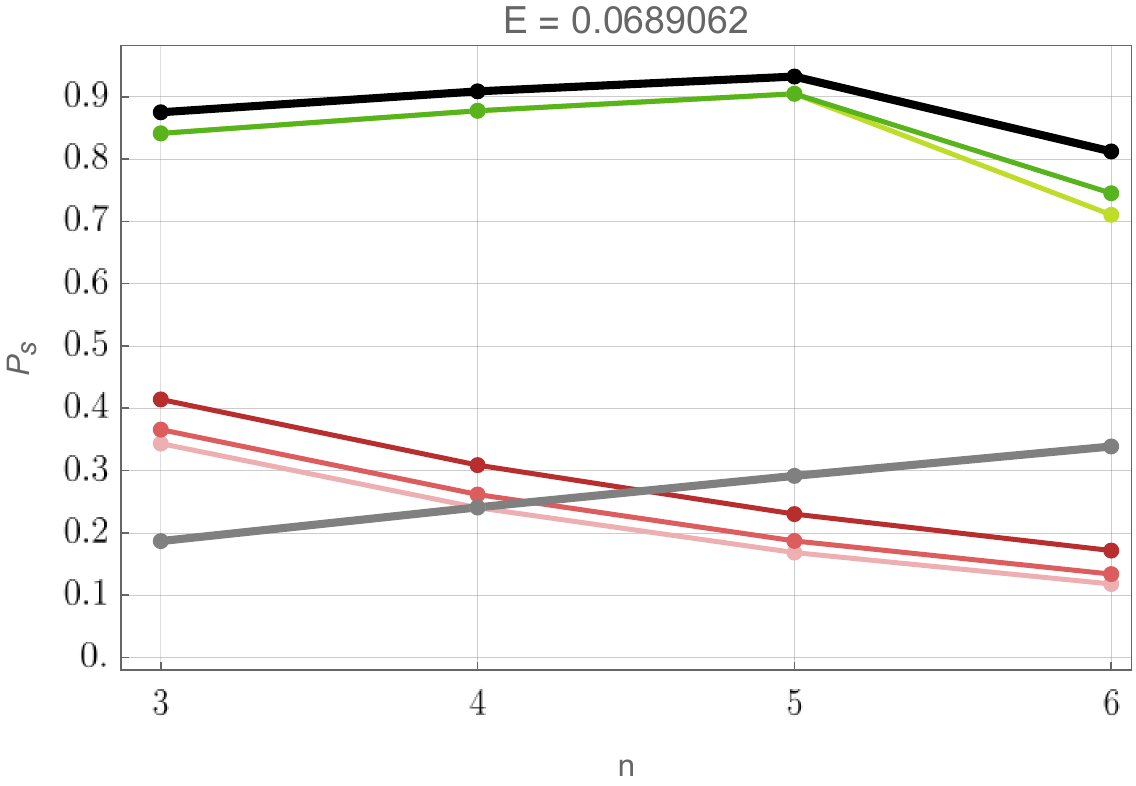} \includegraphics[scale=.32]{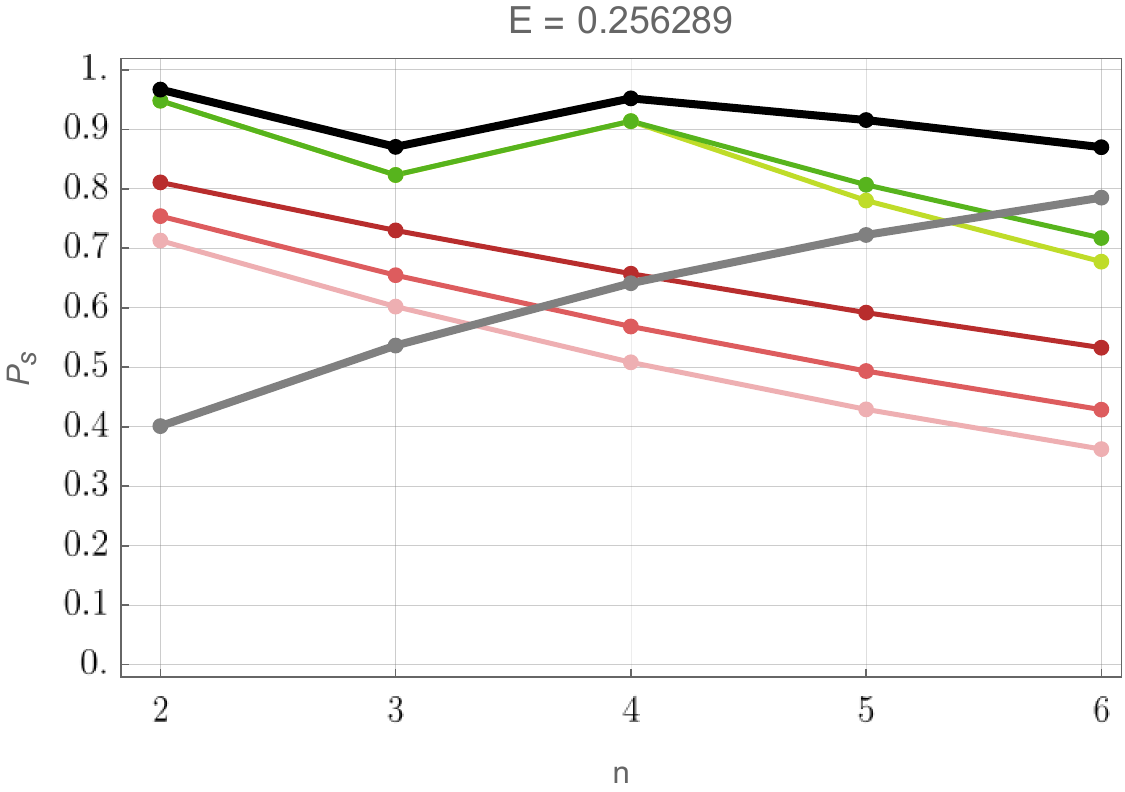}\\
\begin{center}
    \includegraphics[scale=.32]{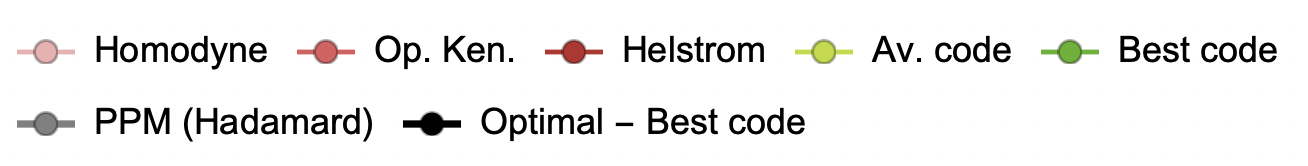}
\end{center}}
\caption{Success probability vs. number of signal modes (message length) for linear codebooks.}\label{fig:psucc_rate_linear_main}
\end{figure}

For linear codes, once the code-size is fixed, several equivalent codebooks $\cC$ are available. Hence, we run different instances of the algorithm, each with a different randomly constructed codebook, identifying for each of them a near-optimal decoder in the chosen circuit class. Then, for each codebook $\cC$ and optimal decoder $U(\btheta_{\rm op}(\cC))$ we analyze its performance by computing its true decoding success probability, i.e., $1-P_{\rm err}(\cC,\btheta_{\rm op}(\cC))$. In Fig.~\ref{fig:psucc_rate_linear_main} we plot the average of this quantity with respect to the analyzed codebooks, as well as its maximum value, corresponding to the codebook and decoder with the largest success probability among those tested, from here on labelled \emph{best codebook}.

The average and maximum decoding success probabilities of our decoders are compared with the following benchmarks:
\begin{itemize}
    \item The theoretically optimal success probability for the best codebook, which is in principle attainable by allowing arbitrary quantum measurements on the collective Hilbert space $\cH$ of the modes, potentially outside the GP class~\cite{Ban1997,Eldar2000,Cariolaro2014,Krovi2014,Pozza2015,Rosati17c,Rengaswamy2020};
    \item The success probability of three schemes employing single-symbol binary-phase modulation and different single-symbol receivers, i.e., homodyne (G)~\cite{serafiniBOOK,Takeoka2008}, optimized Kennedy (GP)~\cite{Takeoka2008,Rosati16a,DiMario2019}, and Dolinar-Helstrom (GPC)~\cite{Dolinar1973,helstromBOOK,Chen2012,Becerra2013,Becerra2013a,Becerra2015,Ferdinand2017,DiMario2022};
    \item The success probability of an $n$-mode pulse-position modulation (PPM) with threshold photodetection on each mode, equivalent to that of a HR for $n$ equal to a power of $2$~\cite{Guha11,Rosati16c}. 
\end{itemize}


The comparison is shown in Fig.~\ref{fig:psucc_rate_linear_main} for two values of the received mean photon number $E$ and as a function of the number of modes $n$. Firstly, we observe that our algorithm is able to find codebooks and decoders that outperform the best single-symbol receivers. Secondly, our codebooks and decoders also outperform the HR for a small number of modes. 
However, the performance of our decoders appears to be decreasing as the number of modes increases, and it can even fall below the HR  for $n\geq 5$; such effect is more pronounced for higher mean photon numbers. This leads to formulate two alternative conjectures: (i) the chosen circuit class is too small to provide a large value of the decoding success probability as the number of modes, and hence the code-size, increases; (ii) the algorithm running on the virtual device requires more computational resources to fully explore the parameter space. Both conjectures can be investigated by extending our algorithm to more complex instances: case (i) can be tackled by expanding the analyzed circuit class with squeezing, ancillae, photon-counting detectors or classical adaptive control, while case (ii) can be tackled by increasing computational resources or running the sampling-based algorithm on a physical device. 

\begin{figure}[h!]
{\centering \includegraphics[scale=.3]{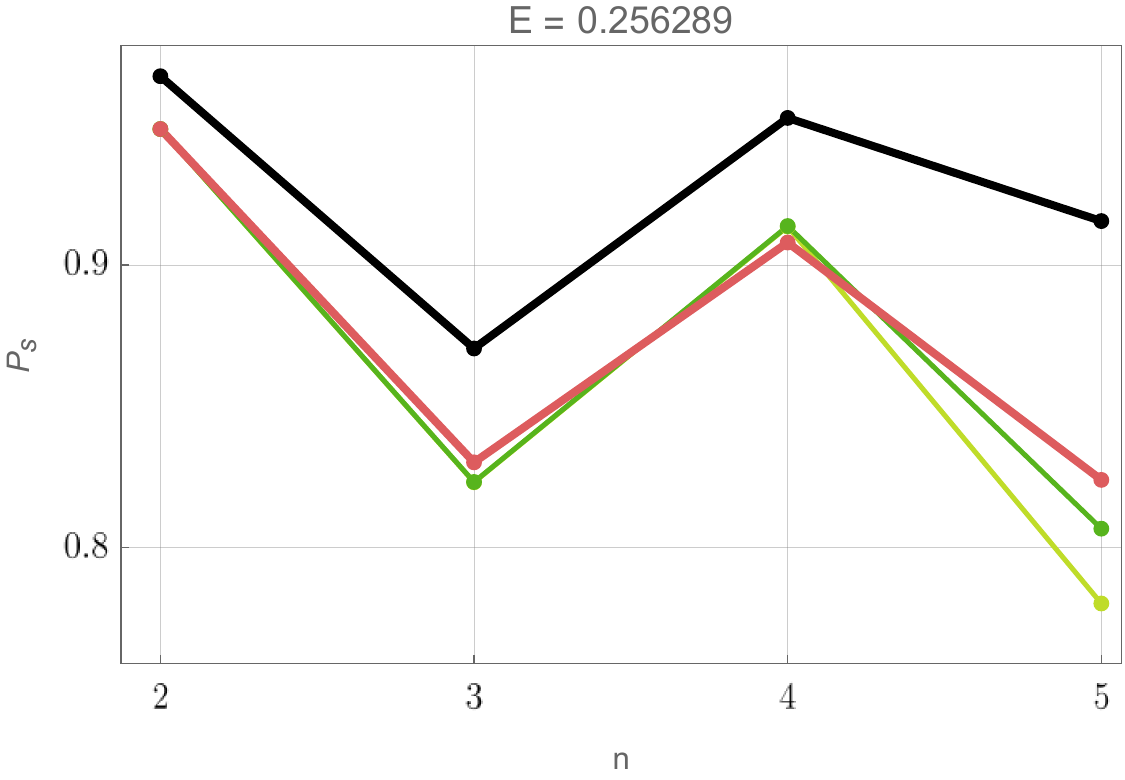} \includegraphics[scale=.3]{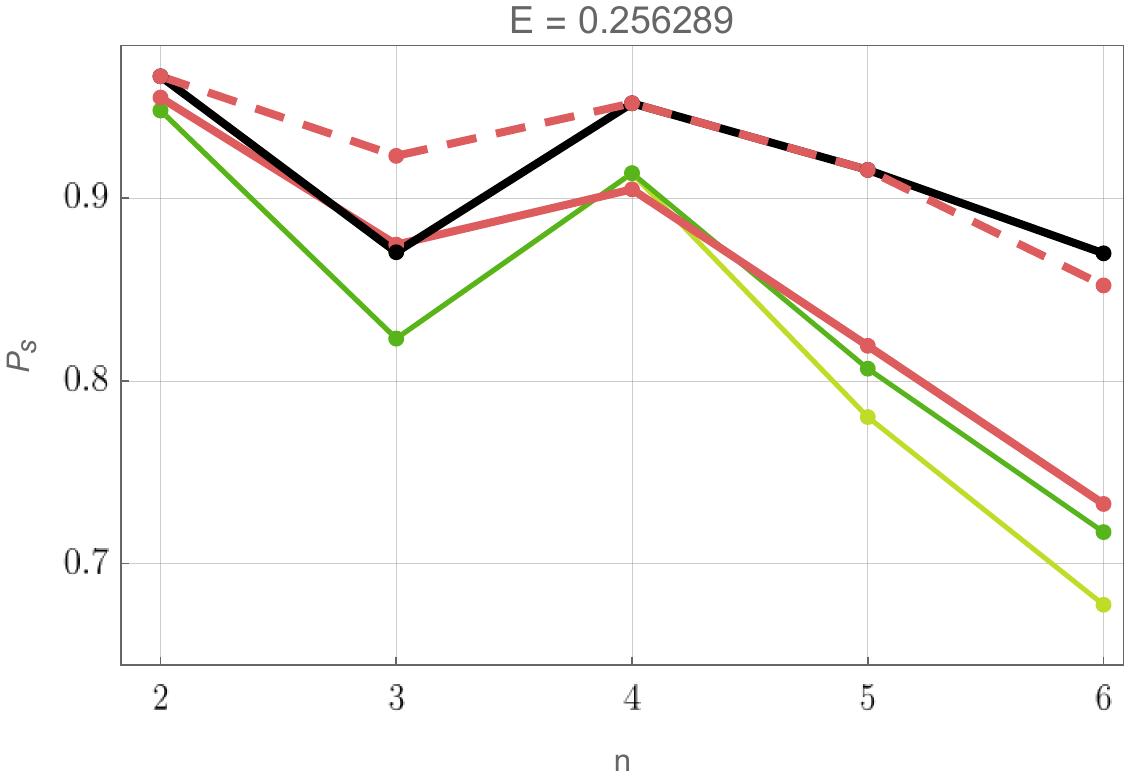}\\
\begin{center}
    \includegraphics[scale=.32]{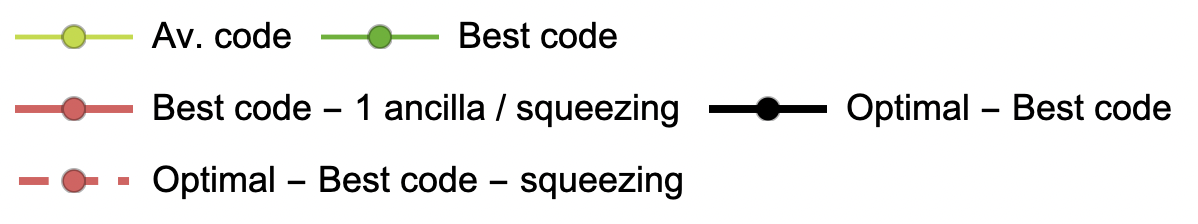}
\end{center}}
\caption{Success probability for linear codebooks using GP decoder circuits with $1$ ancilla (left) or squeezing (right).}\label{fig:psucc_linear_anc_sq}
\end{figure}

Therefore, we further explored the first option (i) by running our algorithm with larger circuit classes, as shown in Fig.~\ref{fig:psucc_linear_anc_sq}. On the left part we considered decoder cirucits of $n+1$ modes, including a fully controllable and measurable ancilla; the results show a very modest advantage, contrarily to what happens for single-mode multi-state unambiguous discrimination~\cite{Sidhu2021}.
On the right we considered decoder circuits of $n$ modes with squeezing operations. In this case, the best codebook and decoders change, resulting in a larger value of the theoretically optimal success probability, consistent with previous works in different contexts~\cite{Takeoka2008,Fanizza2020b}; this increase is partially reflected in the actual success probability with the GP decoder found by the algorithm.
Nevertheless, in both cases it appears that the addition of further resources within the GP class does not bring a significant advantage in the discrimination performance. Therefore, it remains an open problem to explore decoders within the larger GPC class or photon-counting detectors, as well as to improve the training capabilities of the algorithm.

\begin{figure}[h!]
{\centering\includegraphics[scale=.32]{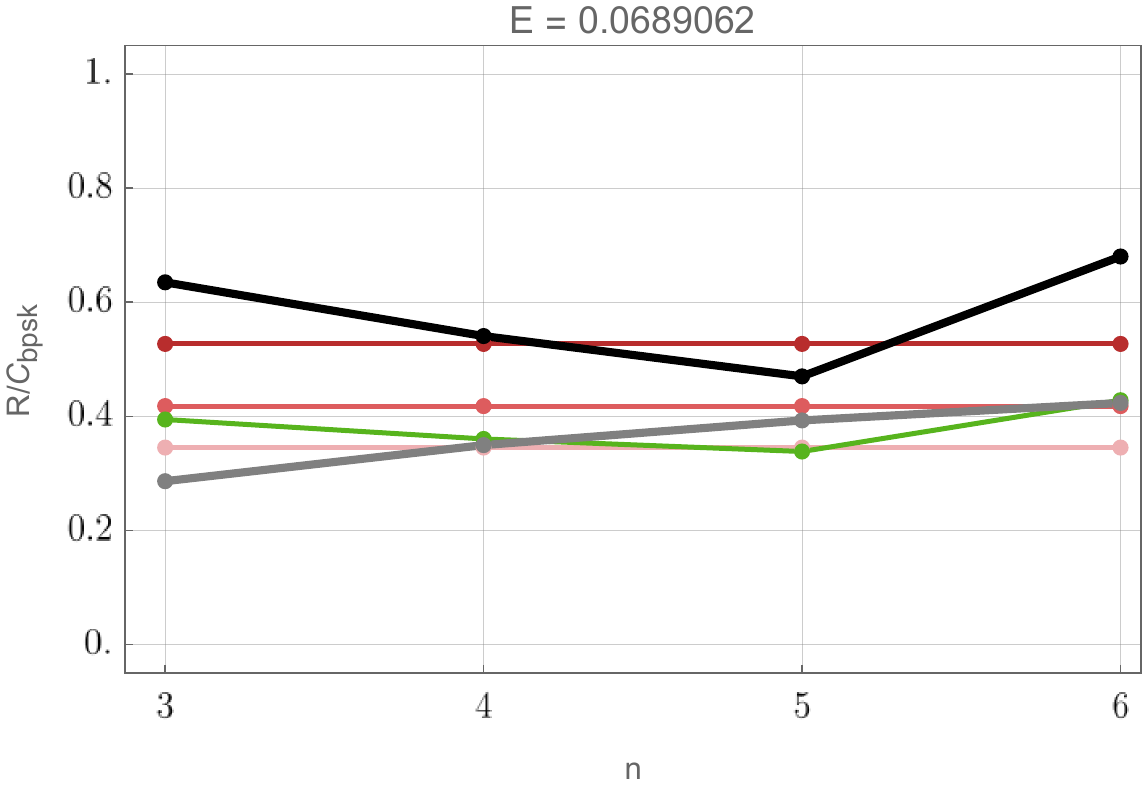} \includegraphics[scale=.32]{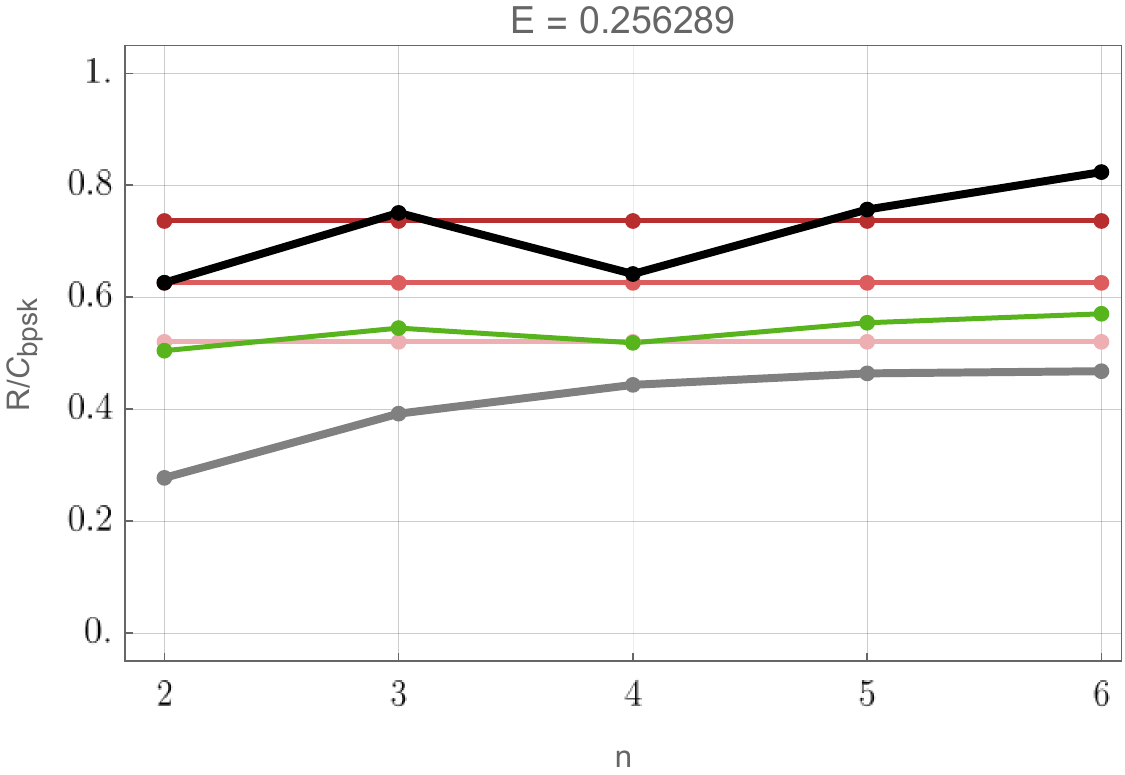}\\
\begin{center}
    \includegraphics[scale=.32]{p_succ_legend.png}
\end{center}}
\caption{Rate per units of binary-channel capacity vs. number of signal modes (message length) for linear codebooks.}\label{fig:rate_linear_main}
\end{figure}

While the decoding success probability is a practically relevant figure of merit, one also needs to evaluate the  bit transmission rate of our discovered codebook and decoders. Let us note that these rates are asymptotic in nature, i.e., they can be attained in the limit of a large number of repetitions of the basic $n$-mode bit-transmission protocol described by the best codebook and decoder. We compare with several quantities:
\begin{itemize}
    \item The capacity of a binary-phase modulation, which attains the Holevo capacity at low received mean photon number but still requires an optimal JDR to be realized in practice~\cite{seqCoh,wildeguha1,Rosati16b,Rosati16c,Rosati2017};
    \item The rate attainable with a binary-phase modulation and various single-symbol receivers, i.e., homodyne, optimized Kennedy and Helstrom~\cite{serafiniBOOK,Takeoka2008,Rosati16c};
    \item  The rate attainable with a $n$-mode PPM with threshold photodetection on each mode, equivalent to that of a HR for $n$ equal to a power of $2$~\cite{Guha11,Rosati16c}. 
\end{itemize}
The results are shown in Fig.~\ref{fig:rate_linear_main} (right column). Remarkably, the decoders found by our algorithm have better than or similar performance to the HR for any photon number and number of modes. Furthermore, at variance with the HR, our decoders surpass the simplest single-symbol technique, i.e., homodyne detection with a binary modulation. Still, more advanced single-symbol schemes based on GP and GPC operations keep outperforming our JDR. Note however that the rate appears to be increasing with the number of modes, suggesting that single-symbol receivers might be surpassed at larger code-size $n>6$.

\begin{figure}[h!]
{\centering\includegraphics[scale=.323]{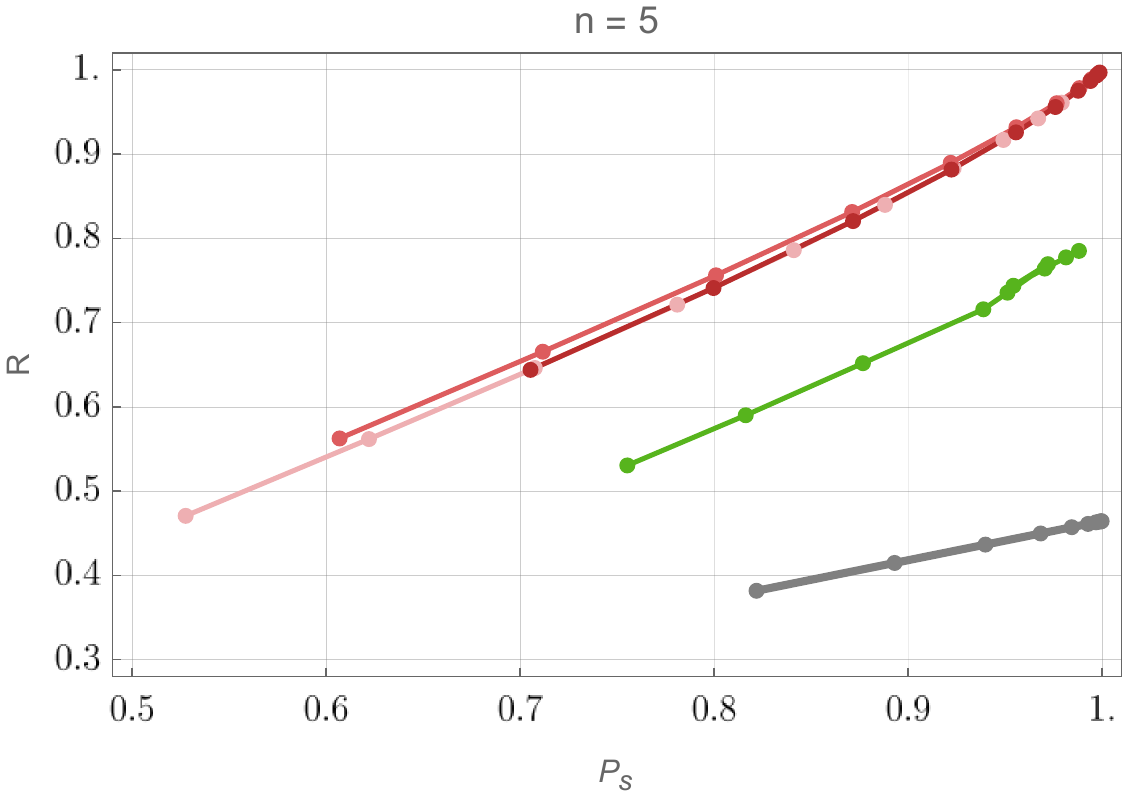}
\includegraphics[scale=.323]{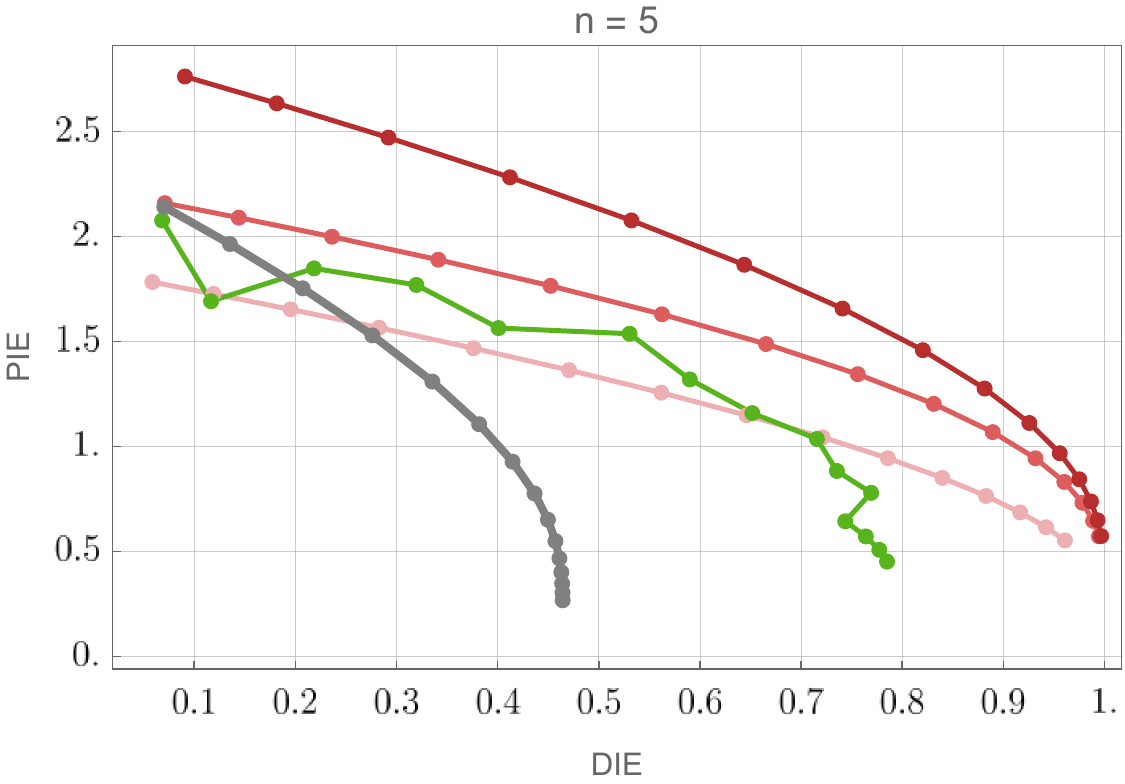} 
\begin{center}
    \includegraphics[scale=.32]{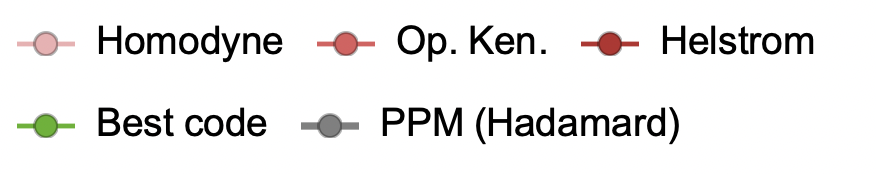}
\end{center}}
\caption{Scatter-plot of rate vs. success probability (left) and photon information efficiency vs. dimensional information efficiency (rate) for linear codebooks, fixing $n=5$ modes and varying the received mean photon numbers $E\in[0.34,1.7]$ (left) and $E\in[0.03,1.7]$ (right). The dots correspond to increasing energies from left to right on the horizontal axis.}\label{fig:scatter_linear}
\end{figure}

The results presented so far were obtained for a relatively small photon number, where the HR can still be considered competitive. In order to analyze our decoders over the whole range of received mean photon numbers studied, up to $E\sim1.7$, in Fig.~\ref{fig:scatter_linear} (left) we further provide a compact representation of our decoders' performance in terms of both rate and success probability, which confirms the previously described features also at larger mean photon numbers. Strikingly, our decoders appear to unlock the access to a whole new region of the parameter space, combining a significantly \emph{larger bit transmission rate than the HR} with a \emph{larger success probability than single-symbol receivers}.

The advantage of our best codebooks and decoders in terms of energy- and mode- efficiency are further encapsulated by Fig.~\ref{fig:scatter_linear} (right): on the horizontal axis, we plot the dimensional information efficiency (DIE), i.e., the number of correctly decoded bits \emph{per mode} sent; instead, on the vertical axis we plot the photon information efficiency (PIE), i.e., the number of correctly decoded bits \emph{per photon} sent. The main limiting factor of the HR is that it provides a large PIE by drastically reducing its DIE, resulting in a seriously inefficient communication protocol as soon as the received mean photon number increases $E\gtrsim 10^{-1}$. Strikingly, this plot demonstrates that our decoders can cure the shortcomings of the HR and surpass the performance of the single-symbol homodyne receiver in an energy range that is practically relevant beyond deep-space communication. 

We conclude our study of near-optimal GP decoders for linear codebooks by examining non-asymptotic bit-transmission rates. Indeed, while the rates shown so far can be attained only in the limit of a large number of repetitions, one can estimate the second-order correction to the asymptotic rate via a Gaussian approximation~\cite{Hayashi2009,Polyanskiy2010,Tan2015,Wilde2016b}. The result is the so-called finite-blocklength rate, i.e., a lower bound on the maximum number of bits transmittable with a certain error threshold $\epsilon$ using only a finite number of repetitions of the basic protocol, called \emph{blocklength}; this quantity gives a more realistic estimate of the rates actually achievable with finite resources. 

\begin{figure}[h!]
{\centering\includegraphics[scale=.325]{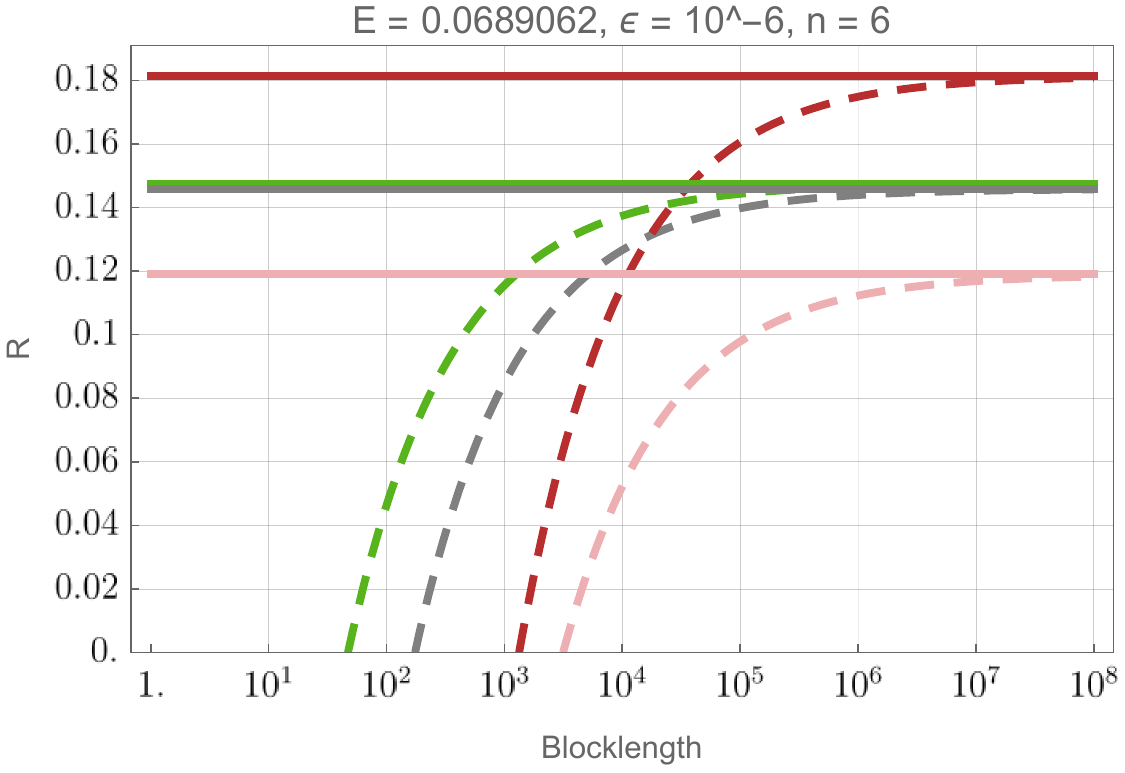} 
\includegraphics[scale=.32]{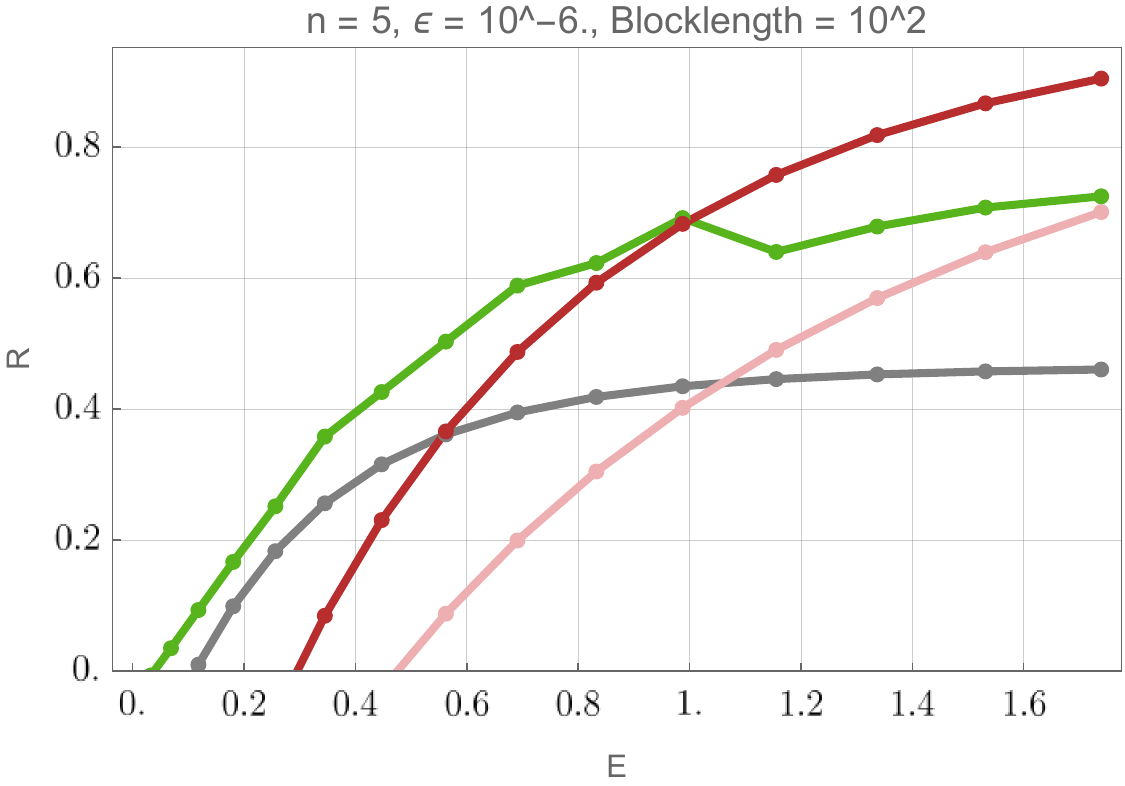} 
\begin{center}
    \includegraphics[scale=.32]{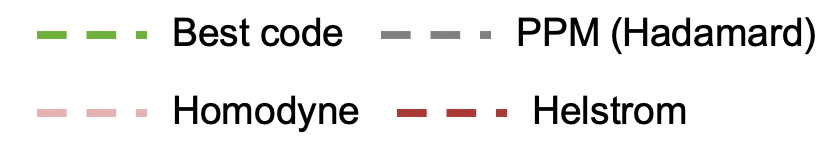}
\end{center}}
\caption{(left) Second-order coding rate vs. blocklength for linear codebooks, at fixed error rate; the solid lines correspond to the asymptotic rates achievable in principle for infinite blocklength. (right) Second-order coding rate vs. received mean photon number for linear codebooks, at fixed error rate and blocklength. }\label{fig:second_order_rate_linear}
\end{figure}

In Fig.~\ref{fig:second_order_rate_linear} (left) we plot the finite-blocklength rate as a function of the blocklength for our best codebook and decoder, as well as for two single-symbol communication protocols and PPM (HR); in Fig.~\ref{fig:second_order_rate_linear} (right) we plot the same quantities fixing the blocklength and varying the received mean photon number. 
These plots show that our discovered decoders \emph{outperform the HR already at small blocklength}, even when the asymptotic rate advantage is minimal. Furthermore, and crucially, they also \emph{significantly outperform the optimal single-symbol modulation protocol}, i.e., binary modulation with Helstrom detection, thus strongly highlighting the practical value of our JDR's in a realistic regime with limited number of channel uses. 

\begin{figure}[h!]
{\centering\includegraphics[scale=.32]{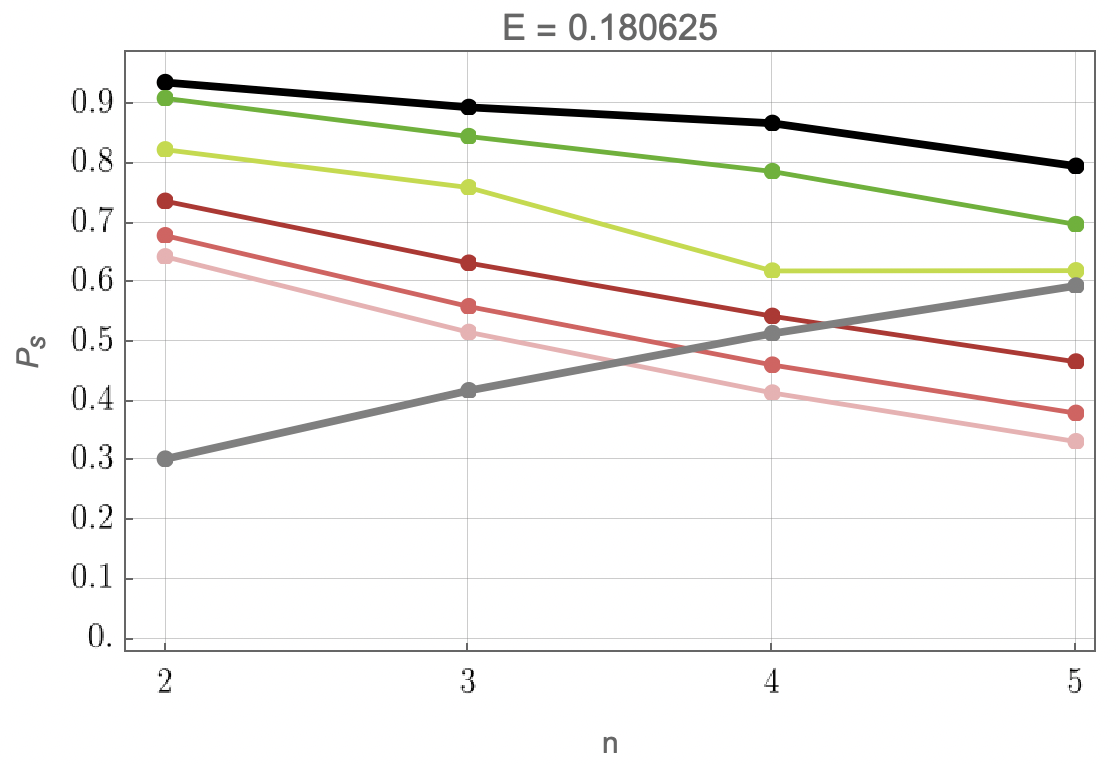} 
\includegraphics[scale=.32]{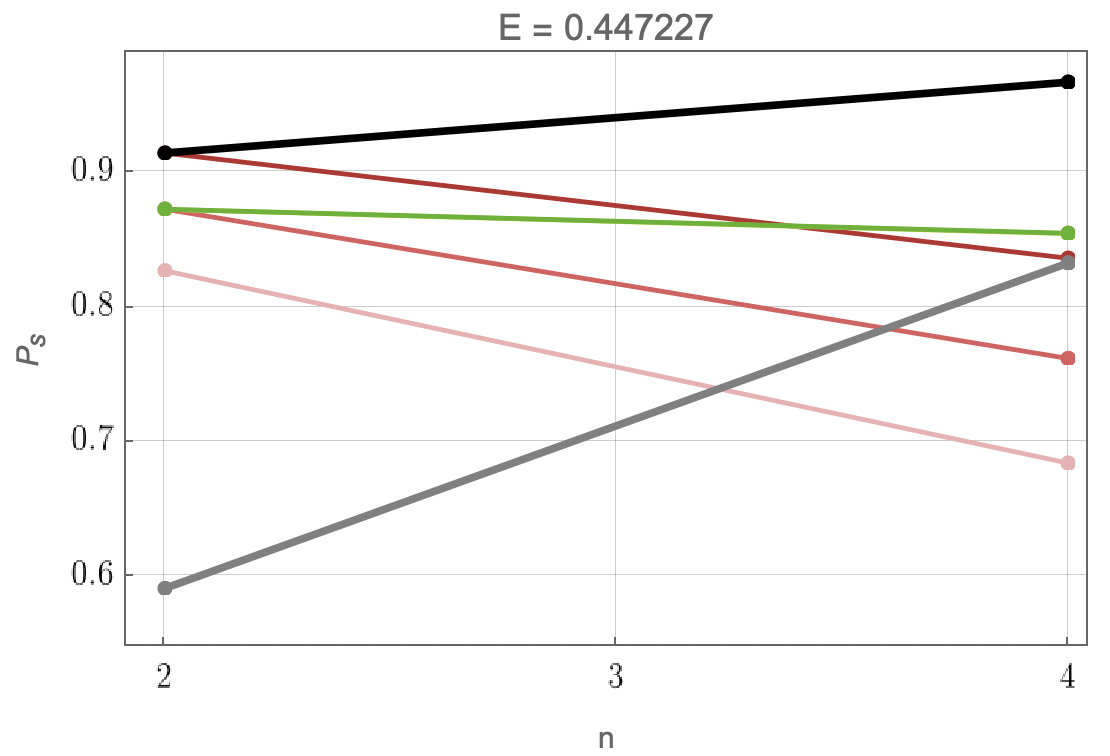} 
\begin{center}
    \includegraphics[scale=.32]{p_succ_legend.png}
\end{center}}
\caption{Success probability vs. number of signal modes (message lenght) for random (left) and polar (right) codebooks.}\label{fig:psucc_random_polar}
\end{figure}

Finally, in Fig.~\ref{fig:psucc_random_polar} we briefly report on the performance of our algorithm with random and polar codes. 
In the former case, as for linear codes, we have to average with respect to all possible random codebooks of a given size, with the added difficulty that there exist many more than for linear codebooks; this limits the efficiency of our algorithm at fixed computational resources, which translates in a quick deterioration of the efficiency in identifying the optimal GP circuit as the energy and the number of modes increases. Nevertheless, the discovered decoders can still outperform competitors for small number of modes and $E<0.25$. 
In the latter case, instead, using the standard polar-code construction of Arikan~\cite{Arikan2009}, generalized to classical-quantum channels by Wilde and Guha~\cite{wildeguha1,Wilde2013}, one is restricted to a number of modes $n$ equal to a power of $2$. Thus we can obtain few data-points, running our algorithm only for $n=2,4$; the results show a discrete advantage with respect to the competitors, and suggest that the receiver can outperform single-mode protocols when increasing the number of modes; it is clear however that a definitive answer can be obtained only by increasing the algorithm's efficiency so to tackle the cases $n=8,16$.\\

\begin{figure}[h!]
{\centering\includegraphics[scale=.4]{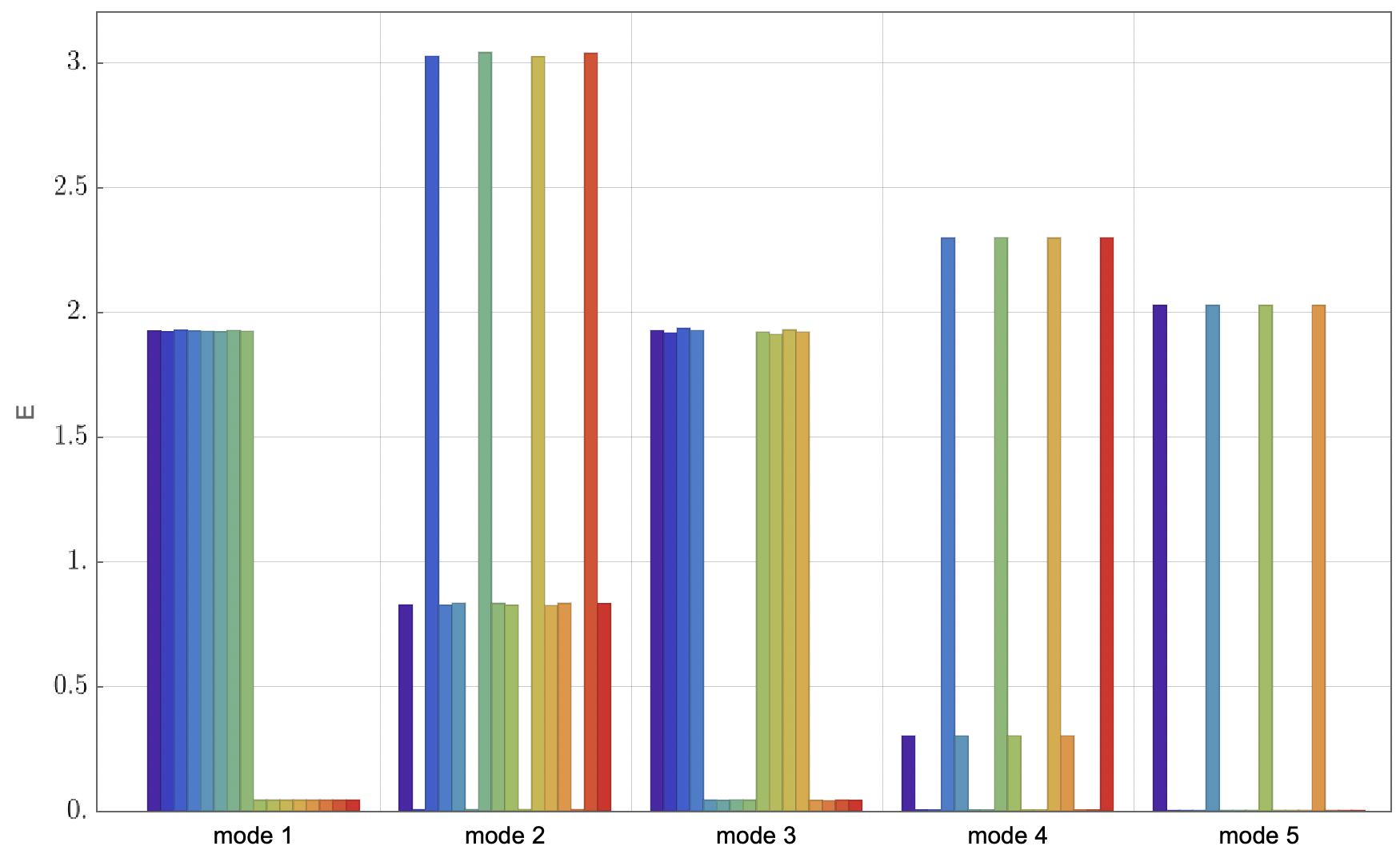}
\begin{center}
    \includegraphics[scale=.32]{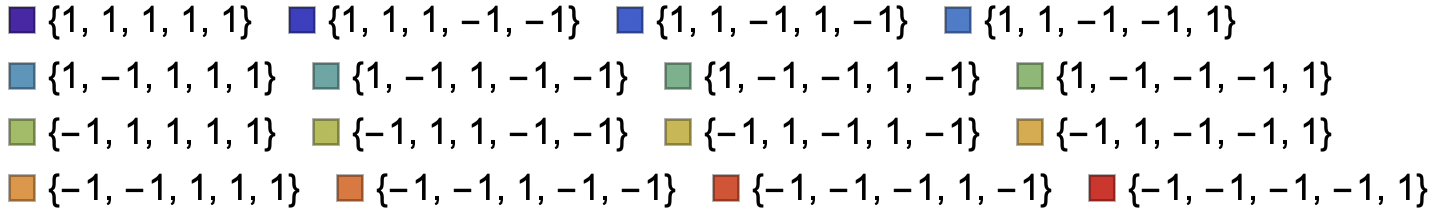}
\end{center}}
\caption{Plot of the mean photon number on each mode for each of the quantum codewords at the output of the GP decoding circuit found by the algorithm. The mean number of photons at the input is $E=0.345156$, the number of modes $n=5$. Different codewords correspond to different colours, as shown by the legend, where each label is given by the sign-modulation of an input codeword. We stress that the output codewords' coherent-state amplitudes are all real-valued, hence a plot of these quantities would have similar features to the present one.}\label{fig:output_codewords}
\end{figure}

We conclude by explicitly analyzing the physical transformation operated by the discovered decoders, restricting to passive linear optics decoders for linear codes. In Fig.~\ref{fig:output_codewords} we plot the mode-by-mode mean-photon-number profile of the quantum codewords at the output of the decoding circuit, just before detection. Indeed, since we are restricting to linear-optical decoders, the unitary $U(\btheta)$ acts on coherent states by rotating their amplitude vector and we can write $U(\btheta)\ket{\balpha} = \ket{\tilde\balpha}$; we thus study the mode-profile, i.e., $|\tilde\alpha_i|^2$ as a function of $i=1,\cdots,n$. We observe very intriguing features: (i) the coherent states that make up the letters of each output codeword have real and positive amplitude, $\tilde\alpha_i(m)>0$ for all $i$ and $m=1,\cdots,|\cC|$, implying that the optimal circuit correctly prepares the codewords for an intensity detection, removing any kind of phase-modulation; (ii) for any two codewords $m,\ell$, there exist one or more modes where they can be discriminated clearly via photo-detection, since their mean-photon numbers on those modes differ by at least one photon, i.e., $|\tilde\alpha_i(m)|^2-|\tilde\alpha_i(\ell)|^2>1$ for one or more values of $i$. 

Upon a careful consideration, these observations are consistent with the intuition behind the HR: in that case, one constructs $n$ specific codewords on $n$ signal modes using a subset of the $2^n$-dimensional set of binary-modulated sequences. The Hadamard circuit simply converts them into PPM, which is optimal for photo-detection measurements, ensuring that each output codeword can be uniquely identified by the mode where it has mean photon number larger than zero. The problem, as noted before, is that the code-size necessary for this property to hold is exponentially smaller than that of a capacity-achieving code.

In contrast, our algorithm identifies much larger codebooks for which more complex photon-allocation rules can be realized by a decoder in the GP class.
These observations lead us to formulate the following conjecture: ``a near-optimal GP decoder for linear coherent-state codes maps the quantum codewords to a multi-level intensity modulation, where each intensity level differs from the others by at least one photon, and different codewords are associated with different intensity profiles". 

\section{Discussion and outlook}\label{sec:discussion}
We have devised and implemented a supervised-learning algorithm for the discovery of classical-quantum decoder designs implementable with current photonic technology. We have extensively analyzed the discovered decoder in terms of numerous performance metrics, based on decoding success probability and bit-transmission rate, and comparing it with the best single-symbol and JDR designs in the literature. 

Our results provide convincing evidence that the algorithm is able to discover original decoder designs surpassing the state-of-the-art. Our decoders unlock the use of non-locality-without-entanglement, a peculiar quantum resource first analyzed by Bennet et al.~\cite{Bennett1999}, for a non-trivial number of parties and for a new range of photon-number values. For the first time to our knowledge, this grants access to a new operative regime characterized by both higher rates and higher decoding success probability, where quantum joint-detection can be deployed in optical-fiber networks.

For example, assuming a loss-per-km of $\alpha=0.05$ and $E_0 = 10^7$ photons at the input (which correspond to $100mW$ input power~\cite{Notzel2022}), our decoders are expected to beat standard homodyne-detection systems at $\sim 350km$, i.e., received photon number $E=0.2511$, providing: (i) a net rate increase for unamplified links (Figs.~\ref{fig:rate_linear_main} and \ref{fig:second_order_rate_linear} right); and (ii) a net reduction of energy consumption for amplified links~\cite{Notzel2022}.\\

From this point onwards, our results open up multiple research directions, including:
\begin{itemize}
    \item The study of generalizations, upgrades or alternatives to the algorithm here employed, and its implementation in a physical photonic device. For example, it might be interesting to substitute the max-likelihood rule with a trainable classical neural network that learns to process the stochastic measurement outcomes; furthermore, an implementation of the algorithm on a physical device might prove quite useful in tackling realistic imperfections via directly including them in the training phase. The ultimate aim here is to tackle larger-scale circuits, larger photon numbers, non-binary modulations and noise-robust designs, all of which can be expected to increase even further the practical advantage of joint detection receivers. 
    \item The realization of practical applications of the proposed designs in fiber networks and free-space links, exploring the above-mentioned possibility of increasing rate and energy efficiency~\cite{Notzel2022}, as well as potential applications in entanglement-assisted communication~\cite{Guha2020}  and continuous-variable quantum cryptography~\cite{Zhang2019,Jain2022a}.
    \item Reaching a fundamental understanding of the circuit structure that unlocks non-locality-without-entanglement in our designs, in order to unveil its physical working principle and, in turn, infer an analytical decoder design for arbitrary code size. 
\end{itemize}

In light of these considerations, we believe that our decoder designs establish the new golden standard for quantum-enhanced bit transmission in the received mean photon number range $E\in[0.1,1]$, ushering an exciting new era for quantum-enhanced communication in near-future networks.

\section{Methods}\label{sec:methods}

\subsection{A discussion of other receivers used for benchmark}\label{subsec:theory_competitors}
A key object to calculate maximum rates and optimal success probabilities is the Gram matrix (see, e.g.,~\cite{Rosati2021}). This is defined, for a codebook $\cC=\{\ket{\psi_i}\}_{i=1}^{|\cC|}$ of quantum codewords, by the square matrix 
\begin{equation}
    \Gamma = \sum_{i,j=1}^{|\cC|}\braket{\psi_i}{\psi_j}\ketbra{\mu_i}{\mu_j},
\end{equation}
and it is positive-definite if $\cC$ is a linearly independent set. In the definition, we used a $|\cC|$-dimensional orthonormal basis $\{\ket{\mu_i}\}_{i=1}^{|\cC|}$ of a virtual Hilbert space. The Gram matrix's central property is it can be directly related to the mean state of the codebook by expressing it in terms of the operator
\begin{equation}
    A = \sum_{i=1}\ketbra{\mu_i}{\psi_i}.
\end{equation}
Indeed, we can write 
\begin{equation}
    \Gamma=A A^\dag,\quad\quad \bar\rho = \frac{1}{|\cC|}\sum_{i=1}^{|\cC|}\ketbra{\psi_i}{\psi_i} = \frac{A^\dag A}{|\cC|}.
\end{equation}
Therefore $\bar \rho$ and $\Gamma$ have the same spectrum, up to a normalization factor. Thus, for example, this can be used to compute the Holevo rate attainable by using a specific codebook $\cC$ and optimal JDR:
\begin{equation}
    \chi(\cC) = S(\bar \rho) = S\left(\Gamma/|\cC|\right).
\end{equation}
Furthermore, it is possible to establish upper and lower bounds on the optimal decoding success probability for the discrimination of the quantum codewords $\cC$ with arbitrary quantum measurements as
\begin{equation}\label{eq:prob_bounds}
  \frac{1}{|\cC|}\sum_{i}(\Gamma^{1/2})_{ii}^2 \leq  p_{\rm succ}^{\rm op}(\cC) \leq \left(\frac{\tr{\Gamma^{1/2}}}{|\cC|}\right)^2 + \sqrt{g_{\max}} \sum_{i}\left|\frac{(\Gamma^{1/2})_{ii}}{\tr{\Gamma^{1/2}}}-\frac{1}{|\cC|}\right|,
\end{equation}
where $g_{\max}$ is the maximum eigenvalue of $\Gamma$, the lower-bound corresponds to employing the square-root measurement (SRM)~\cite{Hausladen1994}, while the upper-bound was derived in~\cite{Sentis2016}. We further stress that the SRM closes the gap \eqref{eq:prob_bounds} and therefore is optimal for all codebooks $\cC$ satisfying specific symmetry conditions~\cite{Ban1997,Eldar2001,Krovi2014,Pozza2015}, e.g., those that can be generated by repeatedly applying a given unitary on one or more seed states, which is the case for linear and polar codes. 

This method can be applied to each of the codebooks analyzed by our algorithm, computing the theoretically optimal bit transmission rate and decoding success probability. These quantities are then used as benchmarks for the performance of the explicit decoders found by our algorithm. \\

The channel capacity \eqref{eq:holevo_capacity} at low received mean photon number $E\lesssim 1$ is well-approximated by the Holevo rate of a binary phase-shift keying (BPSK) modulation $\{|\sqrt E\rangle,|-\sqrt E\rangle\}$:
\begin{equation}
    C_{\rm bpsk}= \chi(\{|\sqrt E\rangle,|-\sqrt E\rangle\}) = H_2\left(\frac{1 - e^{-2 E}}{2}\right),
\end{equation}
where $H_2(p) = -p \log p - (1-p) \log(1-p)$ is the binary-entropy function. This can be attained in the asymptotic limit of multiple communication rounds and by employing an optimal JDR. 

If instead one tries to decode a binary-phase modulation with a single-symbol receiver, the theoretically optimal decoding success probability is given by the Helstrom bound
\begin{equation}
    p_{\rm succ}^{\rm hel}(E)=p_{\rm succ}^{\rm op}(\{|\sqrt E\rangle,|-\sqrt E\rangle\}) = \frac{1+\sqrt{1-e^{-4E}}}{2},
\end{equation}
which is attainable in theory via the corresponding SRM and can be practically approximated to arbitrary precision via the Dolinar receiver~\cite{Dolinar1973,Becerra2013a}, requiring the use of adaptive classical control (GPC class). Simpler single-symbol decoders are given by the generalized Kennedy receiver~\cite{Takeoka2008} (GP class), that uses an optimized displacement operator before detection, with success probability
\begin{equation}
    p_{\rm succ}^{\rm g-ken}(E) = \max_{\beta\in\mathbb{C}}\frac{1+e^{-|\beta-\sqrt E |^2}-e^{-|\beta+\sqrt E |^2}}{2},
\end{equation}
and the standard homodyne receiver~\cite{serafiniBOOK} (G), that performs an optimal measurement of the in-phase component of the incoming field, $x = \frac{a+a^\dag}{\sqrt 2}$, with success probability
\begin{equation}
    p_{\rm succ}^{\rm hom}(E) = \frac{1 + {\rm Erf}(\sqrt{2E})}{2}.
\end{equation}
For these three strategies, the maximum transmission rate can be computed directly as the mutual information of the classical channel mapping the input bit to the output bit after transmission and quantum decoding. Specifically, the induced channel statistics for these three single-symbol communication strategies is given by
\begin{align}
    &p^{\rm hel}(i|i) = p_{\rm succ}^{\rm hel}\; \forall i=0,1,\\
    &p^{\rm g-ken}(0|0) = e^{-|\beta-\sqrt E|^2},\quad p^{\rm g-ken}(1|1) = 1-e^{-|\beta+\sqrt E|^2}, \\
   &p^{\rm hom}(i|i) = p_{\rm succ}^{\rm hom}\; \forall i=0,1,
\end{align}
with $p(j|i)=1-p(i|i)$ for $j\neq i$, and where we have used the optimal $\beta$ for the generalized Kennedy receiver. These translate into the following single-symbol rates
\begin{align}
    & R_{\rm hel}(E) = 1-H_2(p_{\rm succ}^{\rm hel}),\\
    & R_{\rm g-ken}(E) = H_2\left(\frac{p^{g-ken}(0|0)+p^{g-ken}(1|1)}{2}\right),\\
    & R_{\rm hom}(E) = 1-H_2(p_{\rm succ}^{\rm hom}).
\end{align}
Second-order coding rates~\cite{Hayashi2009,Polyanskiy2010} can be computed analogously, based on the general expression
\begin{equation}
    R^{(2)}(\epsilon,b) = R - \sqrt{\frac{2V}{b}} {\rm InverseErfc}(2\epsilon),
\end{equation}
where $\epsilon$ is the error threshold, $b$ the finite blocklength, ${\rm InverseErfc}$ the inverse complementary error function. Furthermore, $V$ is the entropy variance, computable in terms of the channel statistics as 
\begin{equation}
    V = \frac1{|\cC|}\sum_{i,j} p(j|i) \log^2\frac{p(j|i)}{p(j)}. 
\end{equation}

Finally, we consider the only known example of JDR known to date, i.e., the Hadamard protocol~\cite{Guha11,Rosati16c}. This is based on a classical encoding map which is also a linear code, comprising $n$ length-$n$ codewords $\bbb(m)$ given by columns of a Hadamard matrix, i.e., $b_i(m) = H_{m,i}^{(n)}$ and
\begin{equation}
    H^{(n)} = (H^{(n-1)})^{\otimes 2},\quad H^{(2)} = \left(\begin{array}{cc}
       1  & 1 \\
       1  & -1
    \end{array}\right),
\end{equation}
with $n$ a power of $2$.
After encoding and transmission, the quantum codewords are transformed via a passive linear-optical collective circuit (G) into a pulse-position modulation (PPM), where each quantum codeword has a single non-vacuum letter of mean photon number $n$ times larger than at the input. Threshold photodetection and a hard decoding rule are then implemented, inducing the classical channel statistics
\begin{equation}
    p(j|i) = \begin{cases}
      1-e^{-n E}  & j=i, j=1,\cdots,n\\
      0 & j\neq i, j=1,\cdots,n\\
      e^{-n E} & j={\rm err}.
    \end{cases}
\end{equation}
Therefore, the bit transmission rate, entropy variance and success probability of a protocol employing the Hadamard receiver are
\begin{align}
    & R_{\rm hr}(E) = \frac{\log n}{n}(1 - e^{-n E}),\\
    & V_{\rm hr}(E) = \frac{\log^2 n}{n^2} e^{-n E} (2 - e^{-n E}),\\
    & p_{\rm succ}^{\rm hr}(E) = 1 - e^{-n E}.
\end{align}
Finally, we note that the exact same performance can be obtained by employing directly a PPM modulation at any order $n$. In the main text, when $n$ is not a power of $2$ we take the latter protocol as a reference to benchmark our decoders. This can be implemented via a JDR starting from a multi-phase modulation that employs Fourier, rather than Hadamard matrices, i.e., $b_i(m) = F^{(n)}_{m,i} = e^{-\i \frac{2\pi m i}{n}}$.

In conclusion, we stress that the relevance of the HR, or any other GP decoder of the kind studied in the main text, resides in the possibility of starting out with a very symmetric mean-photon-number distribution for the input phase-modulated codewords, and then converting this into intensity-modulated codewords apt for photo-detection. This turns out to be a convenient property, e.g., in the widespread case of power-constrained communication, and it is believed to require the use of non-locality without entanglement or generally the generation of quantum coherence between different symbols~~\cite{Schumacher97,holevo1998c,guha2,Guha2010,Guha11,Giovannetti2011a,Giovannetti2012,guha2012,Chen2012,seqCoh,Takeoka14,Klimek2015,Klimek2015a,Rosati16b,Rosati16c,Rosati2017,Rengaswamy2020,Diaz2020}. If one however is not interested in imposing a photon-number constraint, it is known that, for a wide family of GPC joint-detection schemes, there exists a corresponding output codebook that can be decoded with the same performance via single-symbol measurements~\cite{Rosati2017}.

\subsection{Classical code construction}\label{subsec:codes}
The encoding map $m\mapsto \bbb(m)$ plays a crucial role in preparing the classical message to be later encoded into a sequence of quantum states, and it affects the ultimate performance of the communication protocol. In the classical literature, several coding methods have been studied since Shannon's seminal results.  Here we employ three of them.

Random coding is the method initially employed by Shannon himself to show the achievability of the capacity limit~\cite{Shannon1948}. For a fixed codeword length $n$ and code size $|\cC|$, we construct $|\cC|$ codewords by sampling each letter from the binary alphabet $\{0,1\}$ with equal probability, and considering each codeword only once. The number of possible codebooks constructed in this way is thus given by $\left(\begin{array}{c}
     2^n  \\
      |\cC|
\end{array}\right)$.\\

Linear codes include the most wide-spread families of explicit high-performance codes used nowadays. They are defined by one basic property: a linear combination of two codewords is also a codeword. This linearity constraint is implemented by a certain number of parity checks, which allow to express a given linear codebook in terms of a parity-check matrix or, equivalently, of a generator matrix. For a binary linear code with $k=\log_2 |\cC|$ information bits and $n$ message bits, its generator matrix $G$ in standard form is a $k\times n$ rectangular matrix of the kind
\begin{equation}\label{eq:standard_form}
    G = \left(\one_k | P\right),
\end{equation}
where $\one_k$ is the identity matrix of order $k$ and $P$ a square matrix of order $n-k$ with binary entries. Codewords can be generated by applying $G$ from the right to each $k$-long binary string, obtaining a codebook
\begin{equation}
    \cC(G) = \{\bbb\in\{0,1\}^n : \bbb = \bc G\quad \forall \bc\in\{0,1\}^k\}.
\end{equation}
For a fixed $k$ and $n$, one can generate all linear codebooks in standard form by sampling each entry of $P$ from the binary alphabet with equal probability. Equivalent codebooks, which can be obtained by permutation and linear row operations starting from a $G$ in standard form \eqref{eq:standard_form}, are not taken into account.  The number of possible codebooks constructed in this way is thus given by $2^{k n} = |\cC|^n$. \\

Finally, polar codes~\cite{Arikan2009} are a specific kind of linear codes which exhibit the polarization phenomenon: as the message-length increases, the specific sub-channels mapping the input message to each of the output letters are either noiseless or completely randomizing. This in turn allows for a constructive decoder design. In our case, we follow the standard construction of Arikan~\cite{Arikan2009}: for a given message length $n = 2^\ell$, we employ a generator matrix 
\begin{equation}
    G_n = B_n \left(\begin{array}{cc}
       1  & 0 \\
       1  & 1
    \end{array}\right)^{\otimes \ell},
\end{equation}
where $B_n$ is a permutation known as bit-reversal~\cite{Arikan2009}, which precisely inverts the order of the bits of an input sequence. Then, we compute the mutual information of the so-called split-channels generated by employing the encoded sequences $\bbb = \bc G_n$ for a binary coherent-state modulation, as detailed in the main text. Essentially, each split-channel is identified with one output quantum letter $\ket{\alpha_i}$. Finally, we compute the mutual information $I_i$ of each of these split-channels, employing their symmetric structure (see Sec.~\ref{subsec:theory_competitors}), and select for transmission only a subset of such split-channels satisfying the bound
\begin{equation}
    I_i > \log\left(\frac{2}{1+2^{-\sqrt n}}\right),
\end{equation}
as per~\cite[Theorem 3 and Proposition 1]{wildeguha2}. For the split-channels not selected for transmission, i.e., those which have too small transmission rate, the corresponding bit and quantum letter are frozen to a fixed value before transmission. For example, in the case $E=0.447227$ shown in the main text we have no frozen split-channel when $n=2$ and only one frozen split-channel when $n=4$, corresponding to the first input letter $c_1$ being fixed for all input messages $\bc$. 

\subsection{Circuit components}\label{subsec:circuit}
The circuits we consider for training correspond to the largest class of Gaussian unitary transformations, which can be decomposed in terms of standard optical gates (see, e.g.~\cite{Killoran2018}), amenable to realization via integrated photonics~\cite{Polino2020}: (i) passive linear-optics, i.e., interferometric arrays of beam-splitters and phase-shifters; (ii) optical displacements, which can be realized via a beam-splitter and a strong local oscillator; (ii) single-mode squeezing. The unitary representation of each of these gates is as follows:
\begin{align}
    & U_{\rm bs}(\theta, \phi) = \exp[-\theta (e^{\i \phi} a b^\dag - e^{-\i \phi} a^\dag b)]\\
    & U_{\rm ph}(\varphi) = \exp[-\i \varphi a^\dag a]\\
    & U_{\rm disp}(\beta) = \exp[-\beta a^\dag + \beta^* a]\\
& U_{\rm 1 sq}(r,\phi) = \exp[-r (e^{\i \phi} (a^\dag)^2 - e^{-\i \phi} a^2)],
\end{align}
where all parameters are real-valued, except $\beta\in\mathbb{C}$.

A general $n$-mode passive linear-optical interferometer can be constructed using $n(n-1)/2$ gates $U_{\rm bs}$ and $n-1$ additional $U_{\rm ph}$ gates in several ways; here we employ the rectangular configuration of~\cite{Clements2016}.
In the case where squeezing is absent, the interferometer is simply followed by $n$ optical displacements. Without loss of generality, we can absorb the phases of these displacements into the final phase-gates of the interferometer, thus including an $n$-th phase-gate.
Therefore, the total number of real parameters of an $n$-mode decoding circuit without squeezing and without ancillae is 
\begin{equation}
    \frac{n(n-1)}{2}\cdot 2 + 2n = n(n+1).
\end{equation}

In the case with ancillae, it suffices to increase $n$ by the number of ancillae in the above formula.
In the case with squeezing, two interferometers are employed instead, interleaved by a row of $n$ single-mode squeezing gates, and followed at the end by $n$ displacements. Once again, we can absorb the phases of the squeezing and displacement gates into the interferometers, adding one more phase-gate for each. Therefore, the resulting number of real parameters of the most general $n$-mode decoding Gaussian circuit is
\begin{equation}
    \frac{n(n-1)}{2}\cdot 4 + 4n = 2n(2n+1).
\end{equation}

\subsection{Numerical methods}\label{subsec:algorithms}
Here we describe the steps of the sampling-based supervised-learning algorithm for a physical device.
\begin{enumerate}
    \item Construct a training batch $\cB$ of random codewords with equal probability from $\cC$, and their corresponding classical message label,
    \begin{equation}
    \cB=\{(|\balpha(m_j)\rangle,m_j)\}_{j=1}^{|\cB|};
    \end{equation}
    \item For each codeword $\ket{\balpha(m_j)}$ and fixed parameters $\btheta$, run $N_{\rm shots}$ times the virtual device $U(\btheta)$ with that codeword as input and obtain $N_{\rm shots}$ measurement outcomes $\{\bc_k^{(j)}\}_{k=1}^{N_{\rm shots}}$. Then, estimate the measurement statistics via the relative frequencies of each possible outcome, i.e.,
\begin{equation}
    \hat P(\bc|m_j) = \frac{1}{N_{\rm shots}}\sum_{k=1}^{N_{\rm shots}} \delta_{\bc, \bc_k^{(j)}}.
\end{equation}
    \item After analyzing all the codewords in the batch, for each possible outcome compute the maximum-likelihood guess as
    \begin{equation}
       \hat m(\bc) = \argmax_m P(\bc|m);
    \end{equation}
    \item Compute an estimate of the loss function on the current batch:
        \begin{equation}
            \hat p_{\rm err}(\cB,\btheta) = 1-\frac1{|\cB|}\sum_{\bc} \hat P(\bc | \hat m(\bc)) = 1- \frac1{|\cB| N_{\rm shots}} \sum_{j=1}^{|\cB|}\sum_{k=1}^{N_{\rm shots}} \delta_{\hat m(\bc^{(j)}_k),m_j};
        \end{equation}
    \item Obtain a new candidate set of parameters $\btheta'$ for the next iteration via a function-optimization method.
\end{enumerate}
While this algorithm can be run also using a virtual device by first obtaining the exact expressions of the probabilities and then sampling from them, it is significantly more computationally expensive and we could not use it to obtain systematic results for interesting circuit sizes. We expect that, by running on a physical device that can directly sample outcomes at high speed, this sampling-based version of the algorithm will surpass the performance of the probability-based algorithm described in the main text.\\

For the main probability-based algorithm, we tested several optimization methods and Adam turned out to provide the best results. \\

Theoretical learning guarantees for the above-described algorithms were provided by~\cite{Rosati2022}. In particular, the case of learning with a fixed decision rule using Gaussian plus photodetection circuits is straightforwardly included in Theorem 2 therein, choosing $b\mapsto \alpha (1-2b)$ as the polynomial encoding function for each bit, a fixed coarse-graining of the operators \eqref{eq:measurement_ops} and an ideal target channel that perfectly assigns each input to its corresponding label. Hence, the predicted sample complexity to complete this task up to error $\epsilon$ scales as $O(n^2\epsilon^{-2})$. 

On the other hand, including a maximum-likelihood learning rule requires more careful analysis, since this will cause the association of states and correct outcomes to change depending on the parameter values $\btheta$ that describe the circuit. Nevertheless, for each value of $\btheta$ it is always possible to fix a particular association of states and outcomes corresponding to the one actually realized by the maximum-likelihood rule in that particular training batch. Therefore, including all such state-measurement associations in the concept class provides  a problem formulation that is compatible with~\cite[Theorem 2]{Rosati2022}. Better bounds on the training complexity might be obtained following more advanced approaches to supervised learning with classical-quantum inputs~\cite{Fanizza2022b}.

\section{Data availability}
The optimal codebooks and decoder setups discovered by our algorithm and used for the plots are available as supplementary material.

\section{Code availability}
The code used to obtain our results is publicly available in the Github repository \href{https://github.com/iamtxena/coherent-state-discrimination}{coherent-state-discrimination}.

\section{Author contribution}
MR conceived the research problem, methods and context, designed the training algorithm, identified the targeted codes and parameter regimes, analyzed the resulting data and wrote the manuscript. AS implemented the training, code-construction, circuit simulation and data-processing algorithms.

\section{Competing interests}
The authors declare no competing interests.


\backmatter

\bmhead{Acknowledgments}
MR acknowledges support from the PNRR project 816000-2022-SQUID - CUP F83C22002390007 (Young Researchers). This project has received funding from the European Union Horizon 2020 research and innovation programme under the Marie Sk\l odowska-Curie grant agreement No 845255. MR also acknowledges useful discussions of the results with S. Guha, K. Banaszek and J.-R. Essiambre. 

\end{document}